   \documentclass[preprint]{aastex62}
 \hypersetup{linkcolor=red,citecolor=blue,filecolor=magenta,urlcolor=cyan}
\shorttitle{MANOS II}
\shortauthors{Thirouin et al.}

\begin{document}

 \correspondingauthor{Audrey Thirouin}
\email{thirouin@lowell.edu}

\title{The Mission Accessible Near-Earth Objects Survey: \\
Four years of photometry }

\author[0000-0002-1506-4248]{Audrey Thirouin}
\affil{Lowell Observatory, 1400 West Mars Hill Road, Flagstaff, Arizona 86001, USA.}

\author{Nicholas A. Moskovitz}
\affil{Lowell Observatory, 1400 West Mars Hill Road, Flagstaff, Arizona 86001, USA.} 

\author{Richard P. Binzel}
\affil{Massachusetts Institute of Technology (MIT), 77 Massachusetts Avenue, Cambridge, Massachusetts 02139, USA. }

\author{Eric J. Christensen}
\affil{Lunar and Planetary Laboratory, Department of Planetary Sciences, University
of Arizona, Tucson, Arizona 85721, USA }

\author{Francesca E. DeMeo}
\affil{Massachusetts Institute of Technology (MIT), 77 Massachusetts Avenue, Cambridge, Massachusetts 02139, USA. }

\author{Michael J. Person}
\affil{Massachusetts Institute of Technology (MIT), 77 Massachusetts Avenue, Cambridge, Massachusetts 02139, USA. }

\author{David Polishook}
\affil{Department of Earth and Planetary Science, Weizmann Institute, Herzl St 234, Rehovot, 7610001, Israel. }

\author{Cristina A. Thomas}
\affil{Planetary Science Institute (PSI), 1700 East Fort Lowell Road 106, Tucson, Arizona 85719, USA. }
 
\affil{Northern Arizona University (NAU), San Francisco Street, Flagstaff, Arizona 86001, USA. }

\author{David Trilling}
\affil{Northern Arizona University (NAU), San Francisco Street, Flagstaff, Arizona 86001, USA. }

\author{Mark C. Willman}
\affil{University of Hawaii, Pukalani, Hawaii 96788, USA.}

\author{Brian Burt}
\affil{Lowell Observatory, 1400 West Mars Hill Road, Flagstaff, Arizona 86001, USA.} 

\author{Mary L. Hinkle}
\affil{University of Central Florida (UCF), 4000 Central Florida Blvd, Orlando, Florida 32816, USA. }
\affil{Northern Arizona University (NAU), San Francisco Street, Flagstaff, Arizona 86001, USA. }

\author{Teznie Pugh}
\affil{Lowell Observatory, 1400 West Mars Hill Road, Flagstaff, Arizona 86001, USA.}

\begin{abstract}

Over 4.5 years, the Mission Accessible Near-Earth Object Survey (MANOS) assembled 228 Near-Earth Object (NEO) lightcurves. We report rotational lightcurves for 82 NEOs, constraints on amplitudes and periods for 21 NEOs, lightcurves with no detected variability within the image signal to noise and length of our observing block for 30 NEOs, and 10 tumblers. We uncovered 2 ultra-rapid rotators with periods below 20~s; 2016~MA with a potential rotational periodicity of 18.4~s, and 2017~QG$_{18}$ rotating in 11.9~s, and estimate the fraction of fast/ultra-rapid rotators undetected in our project plus the percentage of NEOs with a moderate/long periodicity undetectable during our typical observing blocks. We summarize the findings of a simple model of synthetic NEOs to infer the object morphologies distribution using the measured distribution of lightcurve amplitudes. This model suggests a uniform distribution of axis ratio can reproduce the observed sample. This suggests that the quantity of spherical NEOs (e.g., Bennu) is almost equivalent to the quantity of highly elongated objects (e.g., Itokawa), a result that can be directly tested thanks to shape models from Doppler delay radar imaging analysis. Finally, we fully characterized 2 NEOs as appropriate targets for a potential robotic/human mission: 2013~YS$_{2}$ and 2014~FA$_{7}$ due to their moderate spin periods and low $\Delta v$.   
 
\end{abstract}

\keywords{minor planets, asteroids: general}

\section{MANOS: Presentation}

Our MANOS project started about 4.5 years ago and aspires to characterize mission accessible NEOs. Our project is designed to fully characterize NEOs, providing rotational lightcurves, visible and/or near-infrared reflectance spectra and astrometry. Such an exhaustive study will give us the opportunity to derive general properties regarding compositions, and rotational characteristics. Because existing physical characterization surveys have primarily centered on the largest NEOs with size above 1~km, MANOS mainly targets sub-km NEOs \citep{Benner2015, Li2015, Reddy2015, Thirouin2016}.  

Our project is split in two main parts: i) \textit{spectroscopy} to provide surface composition, spectral type, taxonomic albedo and infer the object's size, and ii) \textit{photometry} to provide rotational properties and astrometry. Below, we center our attention on the rotational characteristics of the MANOS NEOs extracted from the photometry. 

Here we present new data combined with results from \citet{Thirouin2016}. Thanks to this homogeneous sample of 228 NEOs, we can perform statistical studies and understand the rotational characteristics of the small NEOs in comparison to the larger NEOs. Following, we briefly present our survey strategy and data analysis, in addition to present our lightcurves. Sections 5 and 6 are for our results derived from lightcurves and their implications. Section 7 details our simple model to create three synthetic population of lightcurves assuming different axis ratio distribution for comparison with the literature and our observations. The last section summarizes our conclusions.  

\section{MANOS: Observing plan, facilities, data analysis}

\begin{figure}
\includegraphics[width=15cm, angle=0]{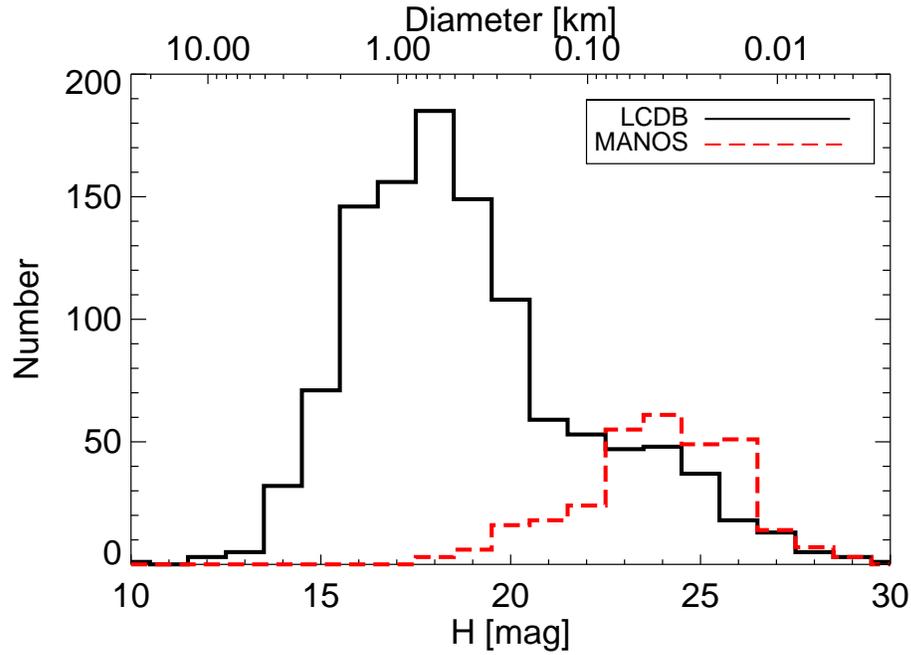}
\caption {The continuous black line summarizes the NEOs from the literature and compiled by \citet{Warner2009} whereas our red discontinuous line represents our MANOS sample observed the past $\sim$4.5~years. We use a ``by-default albedo'' of 20$\%$ to estimate the diameter \citep{Warner2009}.}
\label{Fig:histo}
\end{figure}

In approximately 4.5~years, MANOS observed 308 NEOs for lightcurves (86 objects in \citet{Thirouin2016}, 142 here, and the remainder will be reported in a future work). Figure~\ref{Fig:histo} summarizes the objects observed by MANOS with NEOs from the LCDB\footnote{Lightcurve database (LCDB) from November 2017.} \citep{Warner2009}. The LCDB contains 1,359 entries for NEOs, and 1,147 have a rotation estimate (objects with a constraint for the period are not considered). The LCDB distribution peaks at H$\sim$17~mag (i.e., NEO with a diameter of D$\sim$1~km for a geometric albedo of 20$\%$, \citet{Pravec2007}), whereas for the MANOS sample the peak is at H$\sim$24~mag (i.e., D$\sim$45~m). 

MANOS employs a set of 1 to 4~m telescopes for photometric purposes: 1.3~m Small and Moderate Aperture Research Telescope System (SMARTS) telescope at CTIO, the 2.1~m and the 4~m Mayall telescopes at Kitt Peak Observatory, the 4.1~m Southern Astrophysical Research (SOAR) telescope, and the Lowell's Observatory 4.3~m Discovery Channel Telescope (DCT). A complete description of these facilities, used instruments and filters is available in \citet{Thirouin2016}. In January 2016, Mosaic-1.1 was replaced by Mosaic-3 at the Mayall telescope. This new instrument is also a wide field imager with four 4096$\times$4096 CCDs for a 36$\arcmin$$\times$36$\arcmin$ field of view and 0.26~$\arcsec$/pixel as scale.  

Our observing method and data reduction/analysis is summarized in \citet{Thirouin2016}. Periodograms are in Appendix~A whereas the lightcurves\footnote{Lightcurves and photometry files can be found at \url{manos.lowell.edu}} are in Appendix~B. Typical photometric error bars are $\pm$0.02-0.05~mag, but can be larger in some cases especially with small facilities, faint objects or fast moving objects.

\section{MANOS: Photometry summary}
\label{sec:photo}

For this work, we classified the lightcurves in four main categories: i) \textit{full lightcurve} with a minimum of one entire rotation or a large portion of the lightcurve to estimate a periodicity, ii) \textit{partial lightcurve} showing a decrease or increase of the visual magnitude, but with not enough data for a period estimate, iii) \textit{flat lightcurve} with no obvious increase/decrease in variability and no period detected, and iv) \textit{potential tumblers} with or without the primary period (or shortest period, \citet{Pravec2005}).  
We have full lightcurves for 82 NEOs\footnote{We have 2 lightcurves for 2014~WU$_{200}$. Only the full lightcurve is considered for these estimates.} ($\sim$57$\%$ of our dataset), lower limits for periodicity and amplitude for 21 NEOs ($\sim$15$\%$), flat lightcurves for 30 NEOs ($\sim$21$\%$), and 10 NEOs are potential tumblers ($\sim$7$\%$) (see Figure~\ref{Fig:PeriodSize}). We present two lightcurves (one flat and one full) for 2014~WU$_{200}$. This case will be discussed below.  

MANOS found the fastest known so far rotator: 2017~QG$_{18}$ with a rotation of 11.9~s. This object was imaged at DCT in August 2017, and the lightcurve has a variability of about 0.21~mag. The typical photometry error bar is 0.05~mag. We discovered the potential ultra-rapid rotator: 2016~MA. MANOS observed this object in June 2016, and measured a short period of 18.4~s. The typical photometry error bar is 0.05~mag. The lightcurve displays low variability with a full amplitude of 0.12~mag. Unfortunately, the confidence level of this periodicity is low (i.e., $<$99.9$\%$ confidence level stated for a period estimate) and more data are required to infer if 2016~MA is a ultra-rapid rotator or not. In summary, MANOS discovered four ultra-rapid rotators with periodicities below 20~s: 2014~RC, 2015~SV$_{6}$, 2016~MA, and 2017~QG$_{18}$ (\citet{Thirouin2016}, and this work).

\subsection{Asymmetric/Symmetric and Complex lightcurves}

Only three NEOs display a symmetric lightcurve: 2014~UD$_{57}$, 2014~WF$_{201}$, and 2017~LD, whereas 66 have a bimodal lightcurve with two different peaks \footnote{Tumblers are not considered in this subsection} (i.e. asymmetric curve). 
Majority of the MANOS NEOs has an asymmetry $<$0.2~mag, but sometimes, the difference is higher: 2013~SR and 2015~KQ$_{120}$ with an asymmetry of $\sim$0.5~mag, 2014~FF with $\sim$0.3~mag, 2014~HN$_{178}$ with $\sim$0.4~mag, and 2014~KH$_{39}$ with $\sim$0.7~mag. 

Thirteen objects have complex lightcurves that cannot be fit with only two harmonics: 2014~HS$_{184}$, 2014~HW, 2016~BF$_{1}$, 2016~DK, 2016 ES$_{1}$, 2017~EK, 2017~EZ$_{2}$, 2017~HV$_{3}$, 2017~JM$_{2}$, 2017~KZ$_{27}$, 2017~LE, 2017~MO$_{8}$, and 2017~QX$_{1}$. Reasons fr this morphology are: i) complex shape (NEOs far from spherical/ellipsoidal shapes), and/or ii) albedo contrast, and/or iii) satellite. More observations in different geometries will be useful for shape modeling and to probe for a companion. Unfortunately, most of them won't be brighter than 21~mag in the upcoming decade.  

\subsection{Partial lightcurves}

Twenty-one objects display an increase/decrease in magnitude (red arrows Figure~\ref{Fig:PeriodSize}). We did not calculate a secure periodicity because our observations spanned less than 50$\%$ of the NEO's rotation. For example, 2016~JD$_{18}$ was imaged with Lowell's DCT for a span of $\sim$0.5~h. The partial lightcurve presents a large amplitude of $\sim$1.2~mag and a feature possibly suggesting a complex shape.  

\subsection{Tumblers}

We found ten potential tumblers: 2013~YG, 2014~DJ$_{80}$, 2015~CG, 2015~HB$_{177}$, 2015~LJ, 2016~FA, 2016~RD$_{34}$, 2017~EE$_{3}$, 2017~HU$_{49}$, and 2017~QW$_{1}$. We derive their main periodicities and report them in Table~\ref{Tab:Summary_photo}. For three of them, we are not capable to deduce the main period. In all cases, our data were insufficient to derive the second period with the  \citet{Pravec2005} technique.

\subsection{Flat lightcurves}

Thirty objects have no detected periodicity in the measured photometry. These flat lightcurves can be due to: i) a long/very long periodicity which was not detected over our observing window , ii) a rapid rotation consistent with the exposing time, iii) a (nearly) pole-on configuration, or iv) a NEO with a spheroidal shape. Below, we discuss these four scenarios by assuming that all small NEOs are fast rotators and large NEOs are slow rotator. Such an assumption is based on the well known rotational period-size relation  (Figure~\ref{Fig:PeriodSize}), but it is important to emphasize that our assumption may not be right for all objects as some small objects have been found to be slow rotators \citep{Warner2009}. Thus, MANOS can be identifying slow or fast rotator in the small size range. 

\section{Flat lightcurves: Four scenarios}
\label{sec:flat} 
\subsection{Slow rotators}

As we only dedicate a short observing block per object (typically $\sim$2-3~h, or shorter in case of weather or technical issues), we are biased against long rotational periods (typically, longer than 5-6~h). Five objects from this work and \citet{Thirouin2016} were observed by other teams that derived the following rotation periods: 1994~CJ$_{1}$ ($\sim$30~h, \citet{Warner2015CJ1}), 2008~TZ$_{3}$ (44.2~h, \citet{Warner2009}), 2013~YZ$_{37}$ (8.87~h, \citet{Warner2014}), 2014~SM$_{143}$ (2.9~h, \citet{Warner2015SM143}), and 2015~LK$_{24}$ (18.55~h, \citet{Warner2015}). For 2014~SM$_{143}$, the \citet{Warner2015SM143} observations and ours are separated by about 8-10~days. In both cases, data were obtained at high phase angle ($>$50$^\circ$). We observed 2014~SM$_{143}$ over $\sim$2.5~h with a typical photometric error bar of 0.1~mag and should have detected such a period, assuming that the period derived by \citet{Warner2015SM143} is correct. However, \citet{Warner2015SM143} presented a noisy photometry and their period spectrum showed several solutions that were marginally significant. Therefore, authors not confident about their results, and the reported period could be wrong. Our results about 2014~SM$_{143}$ are available in \citet{Thirouin2016}.

\begin{figure}
\includegraphics[width=18cm, angle=0]{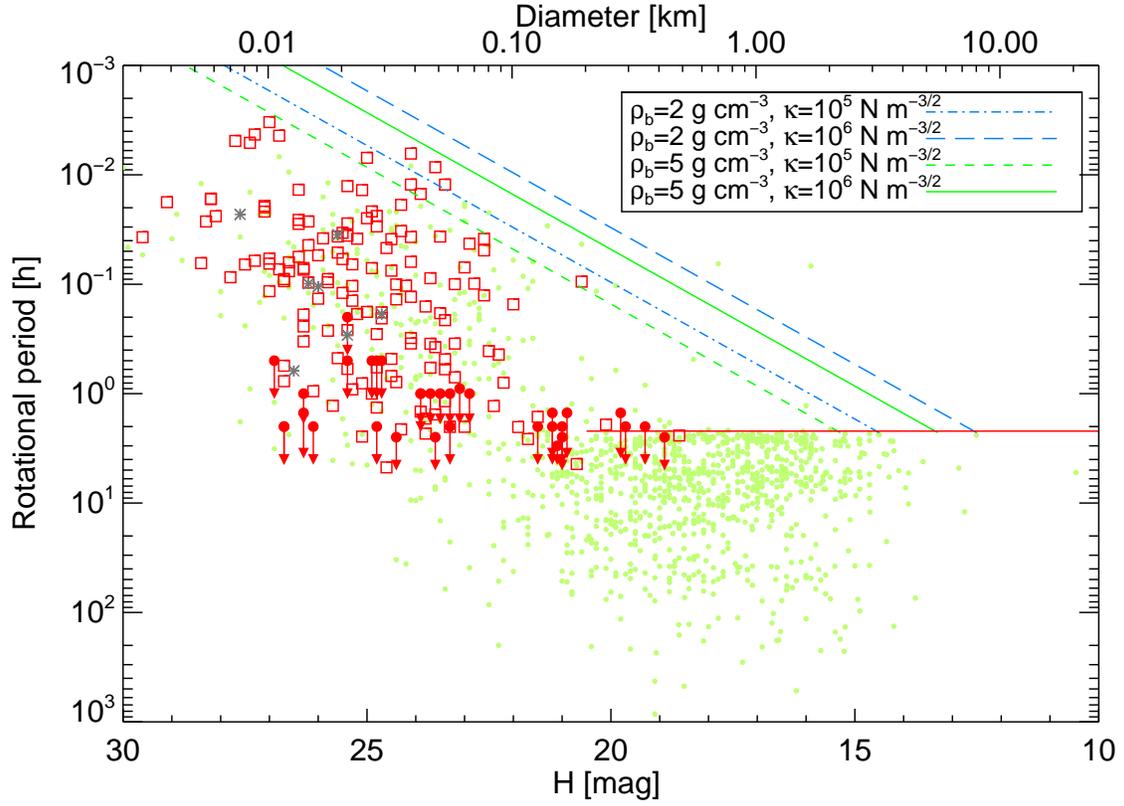}
\caption {MANOS objects with a full lightcurve (red squares), NEOs with a lower limit to their rotations (red arrows), and tumblers (gray asterisks) are plotted. The red continuous line is the spin barrier at $\sim$2.2~h. Blue and green lines are the maximum spin limits assuming different densities and tensile strength coefficients. NEOs from the LCDB are also plotted (green circles).   }
\label{Fig:PeriodSize} 
\end{figure}

We expect ``large'' objects with D$>$100~m (i.e., H$>$22.4~mag) to have a slow rotation (Figure~\ref{Fig:PeriodSize}). Therefore, 2004~BZ$_{74}$, 2005~RO$_{33}$, 2007~CN$_{26}$, 2008~HB$_{38}$, 2010~CF$_{19}$, 2011~ST$_{323}$, 2011~WU$_{95}$, 2012~ER$_{14}$, 2012~XQ$_{93}$, 2014~CP$_{13}$, 2014~OA$_{2}$, and 2014~YD$_{42}$ are probably slow rotators with periods undetected over our short sampling. 2013~UE$_{3}$, and 2016~AU$_{65}$ (H=22.7~mag, and 22.9~mag, respectively) are likely slow rotators too (Figure~\ref{Fig:PeriodSize}). No other published data on these objects for comparison to our results are published. The length of our observing blocks is the lower limit for their periods.  

In conclusion, 19 large objects (D$\gtrsim$100~m) in the full MANOS sample are potential slow rotators (i.e,. $\sim$8$\%$ of the full sample reported in \citet{Thirouin2016} and here). Thus, we estimate that at least 43$\%$ of our flat lightcurves  from this work and our previous paper are caused by slow rotation undetectable over our typical observing blocks. It is crucial to mention that for this estimate, we consider that all large objects are slow rotators which may not be the case for all of them. 

\subsection{Pole-on orientation}

Pole orientations are known for a handful of large NEOs with diameters of several km (e.g., \citet{LaSpina2004, Vokrouhlicky2015, Benner2015}). Shape modeling with radar observations and/or lightcurves obtained at different epochs are required to estimate the pole orientation. MANOS targets typically fade in a matter of hours or days, and their next optical window is often decades away, so lightcurves at different epochs/observing geometries are generally not feasible. For fast and small rotators, radar techniques cannot construct the object's shape, and thus no pole orientation is derived.  

The pole orientation distribution of large objects in the main belt of asteroids (MBAs) is isotropic whereas small MBAs and NEOs (D$<$30~km) have preferentially retrograde/prograde rotation  \citep{Vokrouhlicky2015, Hanus2013, LaSpina2004}. \citet{Vokrouhlicky2015} report 38 pole solutions with an excess of retrograde-rotating NEOs, and noticed a clear deficit of small MBAs and NEOs with a pole orientation of 0$^\circ$. The MANOS set is mostly composed of NEOs in the sub-100m range, and unfortunately, there is no comprehensive information about pole-orientation for this size-range. However, if the sub-100m NEOs follow the same trend as small main belt asteroids and large NEOs, then we expect an excess of small bodies with a pole orientation of $\sim$$\pm$90$^\circ$.

If the rotation axis of an elongated NEO and the sight line are (nearly-) aligned, the brightness variation due to its rotation will be undetectable. Depending of the aspect angle ($\theta$), the lightcurve amplitude of an elongated object (a$>$b$>$c) is:

\begin{equation}
\Delta m = 2.5 \log \sqrt{\frac{\bar{a}^2 \cos ^2 \theta + \bar{a}^2 \bar{c}^2\sin ^2 \theta}{\bar{a}^2 \cos ^2 \theta + \bar{c}^2 \sin ^2 \theta}}
\label{eq:LL03}
\end{equation} 
where $\bar{a}$=a/b, $\bar{b}$=1, and $\bar{c}$=c/b. The likelihood to observe an object pole-on is P~=~1-$\cos$~$\theta$ (\citet{Lacerda2003}). As an example, the probability of viewing a small body with a pole-on orientation $\pm$~5$^\circ$ is $<$1$\%$. Therefore, we estimate that only a few if any of our flat lightcurves are due to a pole-on orientation.  

\subsection{Spherical objects}
\label{sec:angle}

Using the previous equation, the largest amplitude will be at $\theta$=90$^\circ$, and the smallest at $\theta$=0$^\circ$ and 180$^\circ$. At $\theta$=90$^\circ$, $\Delta$m=2.5~$\log$($\bar{a}$). Therefore, the brightness variability of an almost spherical object will be flat. As said, shape modeling using radar observations and/or lightcurves at different epochs are required to derive the object's shape. However, there are very few shape models available for sub-100m NEO \citep{Benner2015}.   

Several NEOs with D$>$200~m have an oblate shape with ridge at the equator or a diamond shape, and they are predicted to be relatively common \citep{Benner2015}. Objects like Bennu, 2008~EV$_{5}$, 2004~DC, 1999~KW$_{4}$, and 1994~CC have an oblate shape based on radar observations, and a low to moderate lightcurve amplitude with periods longer than 2~h \citep{Pravec2006, Ostro2006, Warner2009, Taylor2009, Brozovic2011, Busch2011, Nolan2013, Benner2015}. Assuming that small NEOs are following the same tendency as NEOs with D$>$200~m, some MANOS NEOs are potentially oblate. Oblate objects appear to have long rotational periods that are consistent with/longer than the length of our runs. Therefore, some of our flat lightcurves are potentially caused by oblate objects. Unfortunately, as there is no estimate for the quantity of oblate rotators (independent of size) or if small NEOs have the tendency to be oblate, we cannot propose a clear percentage. 

\subsection{Fast rotators}
\label{sec:ff}

The periodicities of small NEOs (D$<$100m) may be undetected as a result of ``long'' exposure times. For example, we report two lightcurves for 2014~WU$_{200}$. One of the lightcurves is flat, but the second displays periodic photometric variations. The first lightcurve was obtained on November 26$^{th}$ 2014 at DCT. The visual magnitude of 2014~WU$_{200}$ was 20.7~mag (MPC estimate). Due to the faintness and bad atmospheric conditions, we selected an exposing time of 45-55~s (+read-out of 13~s). The typical photometry error bar was 0.03~mag for the DCT data. We re-observed with the Mayall telescope this object few days later when the magnitude was 20.1~mag (MPC estimate). In this case, we employed 10~s as exposure time (+11~s of read-out time), and we favored a rotation of $\sim$64~s. The typical photometry error bar was 0.05~mag for the Mayall data. Therefore, the exposing time used at DCT was too long to derive such a short period. 

Some of our objects with flat lightcurves were imaged with exposing times between 30~s to 300~s. These values were selected for a decent signal to noise, but these times may not have been optimal to sample the lightcurve and so no periodic photometric variations were detected. We estimate that 23 MANOS NEOs are maybe fast to ultra-rapid rotators whose rotation was undetected due to a ``long'' exposing time and/or the bad weather conditions\footnote{Only objects observed with our 4-m class facilities are considered as most of our data are from 4-m class telescopes.}. Small NEOs are commonly rotating fast (Figure~\ref{Fig:PeriodSize}), and if so, 52$\%$ of our flat lightcurves from this work and \citet{Thirouin2016} are potentially due to small ultra-rapid/fast rotators. 

For fast/ultra-rapid NEOs rotating in few seconds or few minutes, the exposure time is important. Following \citet{Pravec2000b}, the optimum exposure time (T$_{exp}^{opt}$) to detect a lightcurve with two harmonics is: 
\begin{equation}
T_{exp}^{opt} = 0.185 \times P
\label{eq:eq3}
\end{equation}
with P as the object's periodicity (Section 2 of \citet{Pravec2000b}). This relation is based on theory and does not reflect a specific observing strategy. Because we know the exposing time during our observations, we can figure out the detectable rotational period. For example, with T$_{exp}$=11~s, we will perfectly sample the lightcurve of a small body rotating in 1~min or more. In this case, an object rotating in $\ll$1~min will have a flat lightcurve and thus its rotation will be undetectable. 

\begin{figure}
\includegraphics[width=18cm, angle=0]{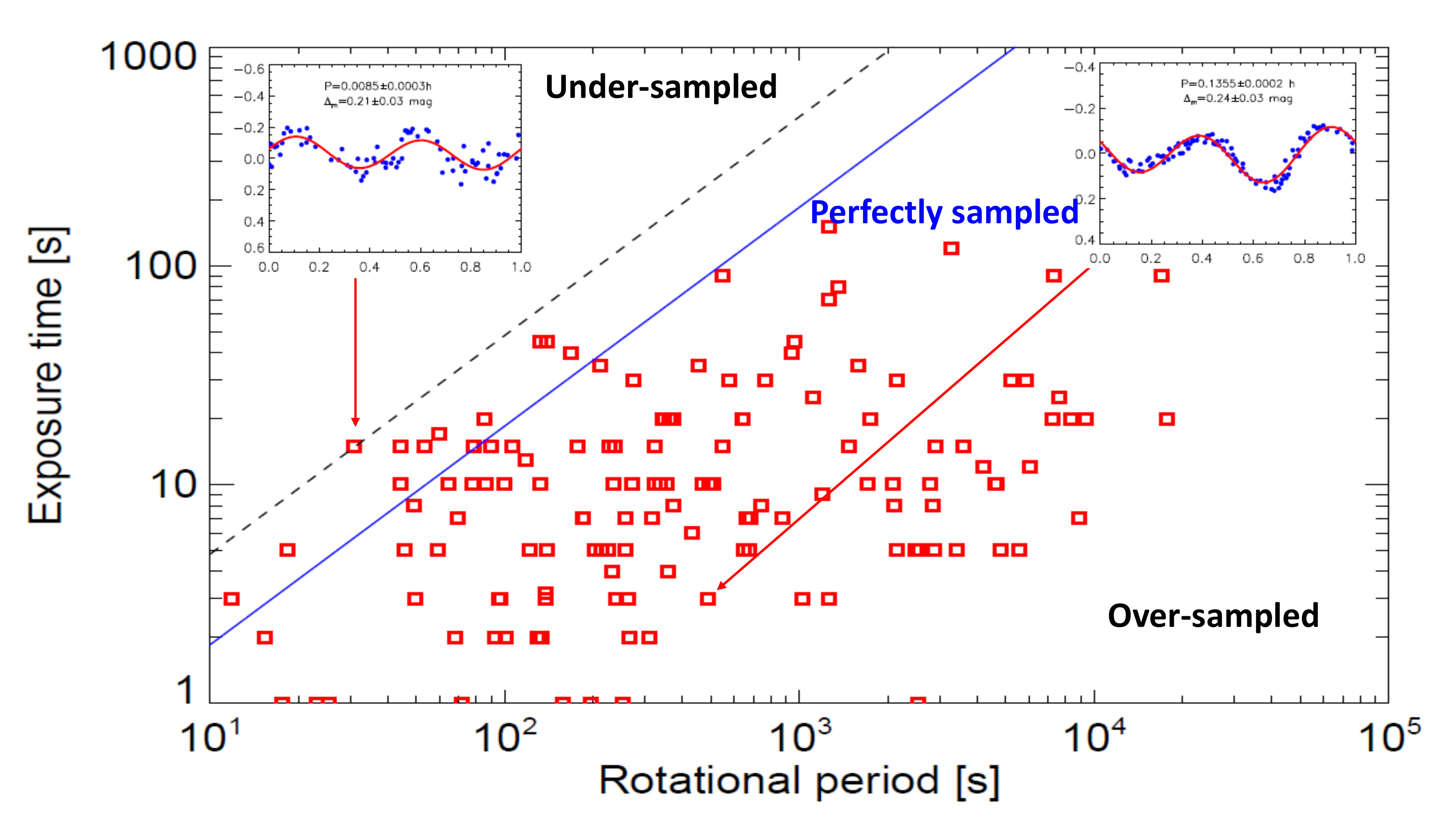}
\caption {Red squares are the MANOS objects with a rotational period estimate. The blue continuous line indicates the relation between exposure time and rotational period for a perfectly sampled lightcurve with two harmonics. Objects below this line have an over-sampled lightcurve, and objects above it have an under-sampled lightcurve. Some MANOS objects have an under-sampled lightcurve, but we were able to derive their rotational period. See Section~\ref{sec:flat} for more details.   }
\label{Fig:exp} 
\end{figure}

In Figure~\ref{Fig:exp}, the continuous line is for Equation~\ref{eq:eq3} for a perfectly sampled two harmonic lightcurve. Data points are MANOS NEOs imaged with our 4-m facilities. Objects below the continuous line have over-sampled lightcurves whereas above this line the lightcurves are under-sampled. The dash line in Figure~\ref{Fig:exp} represents an empirical upper limit to the period-exposure time relationship using the MANOS dataset and can be articulated as: 
\begin{equation}
T_{exp}^{MANOS} =0.48 \times P 
\label{eq:eq4}
\end{equation} 
This relation would converge to Nyquist sampling theory in a regime of infinite signal-to-noise-ratio. For the smallest objects, and thus potentially fast to ultra-rapid rotators, using Equation~\ref{eq:eq4} we can identify the rotational period to which we were sensitive based on object-specific exposure time. Using Equation~\ref{eq:eq3}, and Equation~\ref{eq:eq4}, we have two lower limits for the potential rotational periods.
Therefore, if these objects have a rotational period between these two estimates, we should have detected it. In conclusion, the rotational period is likely shorter than the estimate and thus we undetected it in our observing block (assuming that the objects have a two harmonics lightcurve). But, it is also important to emphasize that some small objects (sub-100m objects), even if they are expected to rotate fast, some might be slow rotators (Figure~\ref{Fig:PeriodSize}).

\section{Physical Constraints}
\label{sec:axis}

A strenghtless rubble-pile will not be able to rotate faster than about 2.2~h without breaking up \citep{Pravec2002}. But, most small NEOs have rotational periods of a few seconds or minutes. Therefore, an explanation of these rapid rotations is that NEOs are bound with tensile strength and/or cohesive instead of just gravity. Using \citet{Holsapple2004, Holsapple2007}, we calculated the maximum spin limits assuming different densities and tensile strength coefficients for the NEO population. Following \citet{Richardson2005}, we considered a friction angle of 40$^\circ$, and moderately elongated ellipsoids (c/a=b/a=0.7). We used two values for the density; 2 (Itokawa \citep{Fujiwara2006}) and 5~g cm$^{-3}$ (density of a stony-iron object \citep{Carry2012}), and two tensile strength coefficients, 10$^5$, and 10$^6$~N~m$^{-3/2}$ (range of tensile strengths for Almahata Sitta, \citet{Kwiatkowski2010}). Five MANOS targets require a tensile strength coefficient between 10$^{5}$-10$^{6}$~N~m$^{-3/2}$: 2014~FR$_{52}$, 2014~PR$_{62}$, 2015~RF$_{36}$, 2016~AD$_{166}$, and 2016~AO$_{131}$.  \\
 
The lightcurve amplitudes ($\Delta m (\alpha)$) in Table~\ref{Tab:Summary_photo} were obtained at a phase angle, $\alpha$. At $\alpha$=0$^\circ$, the amplitude is: 
\begin{equation}
\Delta m (\alpha = 0^\circ) = \frac{\Delta m (\alpha)}{1 + s \alpha }
\label{eq:alpha}
\end{equation}
with s=0.03~mag deg$^{-1}$ \citep{Zappala1990}. In the MANOS sample, only 12 objects (10$\%$ of our sample) have a $\Delta m (\alpha = 0^\circ)$$\geq$0.5~mag, and one object has a $\Delta m (\alpha = 0^\circ)$$\geq$1~mag. In the LCDB, there are 309 NEOs\footnote{Observing circumstances or lightcurve amplitude are not reported for some LCDB objects, and thus they are not considered here. Only NEOs with a H$\geq$20~mag are considered because MANOS focuses on small objects. We select objects observed at a phase angle lower than 100$^\circ$ because MANOS is observing in that range.} with an absolute magnitude H$\geq$20~mag, and observed at a phase angle $\alpha$$\leq$100$^\circ$: 47 of them have a $\Delta m (\alpha = 0^\circ)$$\geq$0.5~mag (15$\%$ of the LCDB), and 6 have a $\Delta m (\alpha = 0^\circ)$$\geq$1~mag (2$\%$). Therefore, the relative abundance of high amplitude lightcurves in these two data sets are consistent.\\

\section{Potential mission targets}

One of our goals is to find favorable target(s) for a future mission to a NEO, and thus mission accessibility is one of our selection criteria \citep{Abell2009, Hestroffer2017, Bambach2018}. For this purpose, we estimate the velocity change for a Hohmann transfer orbit also known as $\Delta v$. A rough guess of the $\Delta v$ is estimated with the \citet{Shoemaker1978} protocol ($\Delta v$$^{SH}$). In order to obtain an accurate estimate, one can use the Near-Earth Object Human Space Flight Accessible Targets Study (NHATS) orbital integration, $\Delta v$$^{NHATS}$ \footnote{\url{http://neo.jpl.nasa.gov/nhats/}}. NHATS uses specific constraints to compute the $\Delta v$$^{NHATS}$: i) Launch before 2040, ii) Total mission duration $\leq$450~days, and iii) Number of days spent at the object $\geq$8~days. The NHATS limit is $\Delta v $$^{NHATS}$ of 12~km~s$^{-1}$. Several of our targets are not following these criteria and so, no $\Delta v$$^{NHATS}$ are available for them (Table~\ref{Tab:Summary_photo}). 

According to NHATS, 78 MANOS NEOs are accessible by a spacecraft (Table~\ref{Tab:candidates}, and Table 2 in \citet{Thirouin2016}). For diverse reasons, \citet{Abell2009} consider that the best target for a mission should have a moderate to slow rotation (P$>$1~h). Only 9 MANOS NEOs have such a long rotation, have a $\Delta v$$^{NHATS}$$\leq$ 12~km~s$^{-1}$; and have been observed for spectroscopy (Table~\ref{Tab:candidates}, and Table 2 in \citet{Thirouin2016}). We will present spectral results for these objects in future publication(s).

Finally, we note that several non-fully characterized MANOS NEOs have a new optical window in the upcoming years or decades. For example, the low $\Delta v$$^{NHATS}$ and slow rotator 2013~XX$_{8}$ (spectral type unknown) will have a new optical window in 04/2019 and thus we will have an opportunity to fully characterize this potential target. 

\section{MANOS+LCDB versus synthetic population}
\label{sec:synthetic}

In this section, we aim to compare our results to a synthetic population of NEOs to identify biases regarding our measured amplitude distribution and to constrain the distribution of morphologies in the NEO population. In a first step, we create 10,000 synthetic objects and calculate their lightcurve amplitude versus aspect angle. In a second step, we ``observationally sample'' this synthetic population based on prescribed phase angles, in order to compare our synthetic population with the MANOS+LCDB data set. 

\paragraph{Step 1:}

\begin{figure*}
\includegraphics[width=11.2cm, angle=0]{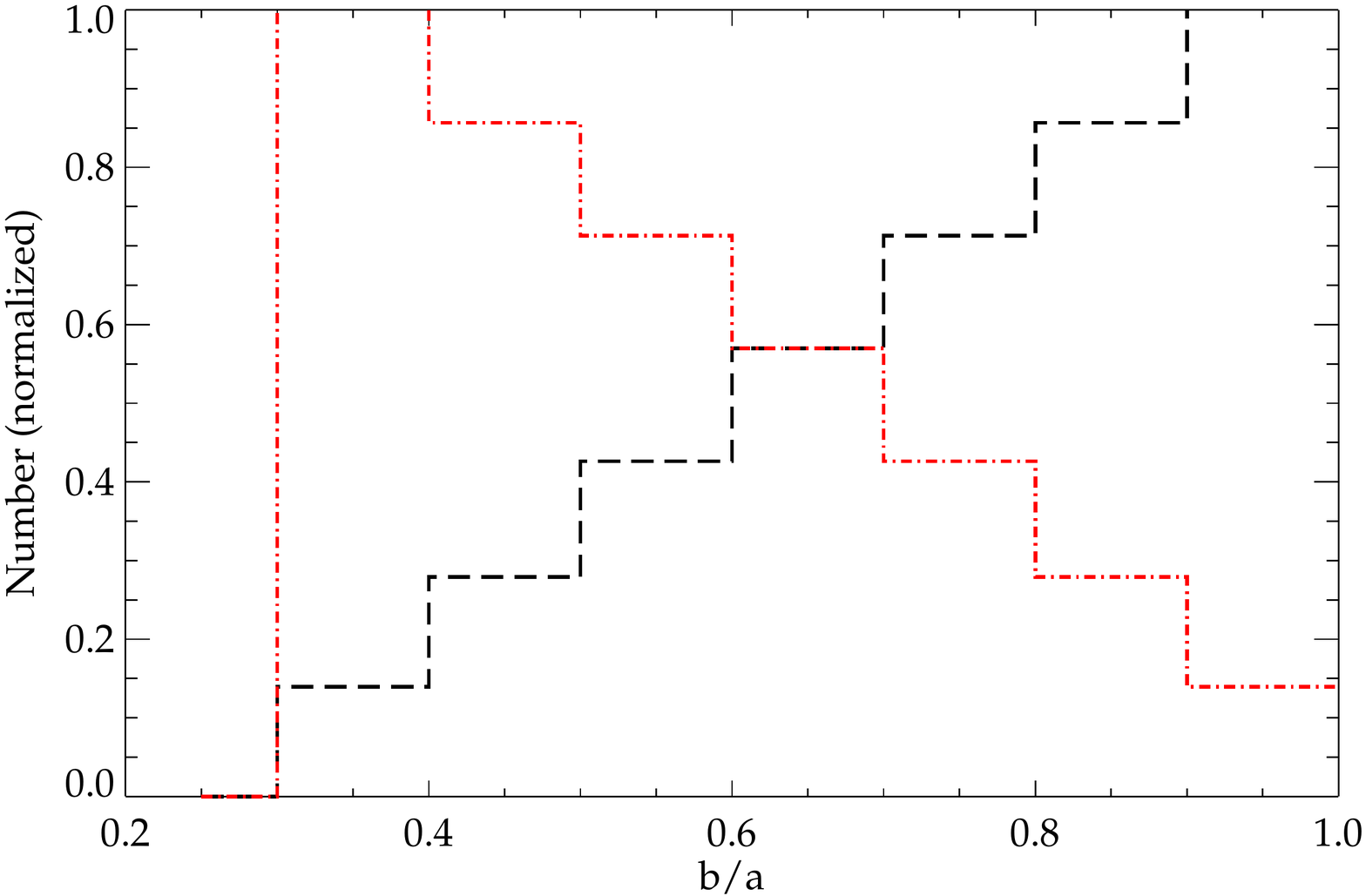}
\\
\includegraphics[width=11.2cm, angle=0]{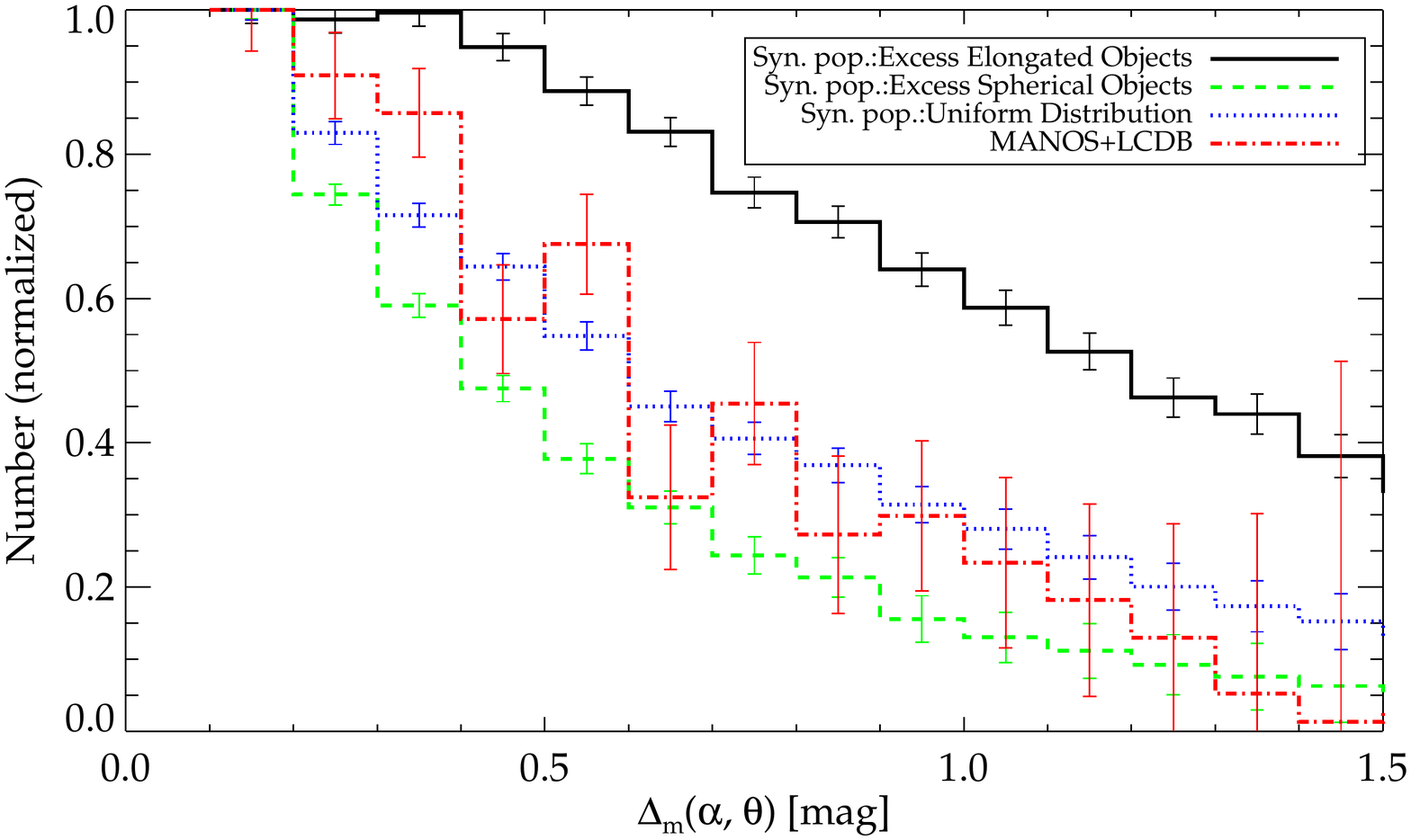}
\caption {We consider two non-uniform distributions of b/a (upper panel) with an excess of elongated or an excess of spheroidal objects. Following the procedure presented in Section~\ref{sec:synthetic} the other plots report the lightcurve amplitude non-corrected from phase angle of the two synthetic distributions, as well as the MANOS+LCDB sample. The lower panel is focusing on objects with an amplitude between 0.1 and 1.5~mag. Error bars are $\sqrt{1/N}$ with N being the number of objects per bin.  }
\label{Fig:excess}
\end{figure*}

Assuming that NEOs are prolate ellipsoids (with b=c) at a phase angle of 0$^\circ$, the amplitude varies as:
\begin{equation}
\Delta m (\theta) = 1.25 \times \log \Big[ \frac{1}{\cos ^2 \theta + (b/a)^2 \sin ^2 \theta} \Big] 
\label{eq:theta}
\end{equation}
where $\theta$ is the aspect angle, and b/a is the elongation of the object \citep{Michalowski1990}. The aspect angle is: 
\begin{equation}
\cos \theta = -\sin \beta_g \sin \beta_p -\cos \beta_g \cos \beta_p \cos(\lambda_g - \lambda_p)
\label{eq:theta2}
\end{equation} 
where $\beta_p$ and $\lambda_p$ are the object's north pole ecliptic latitude and longitude, and $\lambda_g$ and $\beta_g$ are the object geocentric ecliptic coordinates  \citep{Michalowski1990}. 
We use Equation~\ref{eq:theta} to generate the lightcurve amplitude of 10,000 synthetic objects. The only two free parameters in this equation are the axis ratio b/a and the viewing angle $\theta$. In theory, the axis ratio b/a varies from 0 to 1. However, for objects visited by spacecraft, Eros\footnote{Only objects visited by spacecraft were taken into account because of the direct estimate of their size/axis ratio.} is the most elongated with a ratio b/a=0.32 \citep{Veverka2000}. Thus, we limit the axis ratio b/a between 0.32 and 1 (spherical object). We considered three possible axis distributions for our synthetic population: i) a uniform distribution of b/a, ii) one distribution with an excess of spheroidal objects and iii) one with an excess of elongated objects (Figure~\ref{Fig:excess}, upper panel). 

The second parameter is the aspect angle $\theta$ ranging from from 0$^\circ$ to 90$^\circ$ (absolute value). \citet{LaSpina2004, Vokrouhlicky2015} noticed an excess of retrograde-rotating NEOs (based on a limited sample) which would imply that the observed distribution of pole orientations is not uniform. We updated the distribution of poles reported in \citet{Vokrouhlicky2015} with newest results from the LCDB (multiple systems have been excluded from the distribution as we do not expect any small NEO as binary/multiple, \citet{Margot2002}). With the newest results, the pole distribution is still consistent with the \citet{Vokrouhlicky2015} result. Using our updated pole distribution, we created a non-uniform distribution of pole orientation and thus a distribution of ($\lambda_p$;$\beta_p$). Even though most of the objects with a known pole orientation are large objects, and we assume that the pole orientation of the small objects is similar that of large objects. This assumption might be wrong and will need to be tested once more pole orientations of small objects are known. The typical uncertainty on pole orientation is about $\pm$10$^\circ$ based on radar and lightcurve inversion results, so we estimated the number of objects within a grid of 10$^\circ$$\times$10$^\circ$. We use the number density of objects in this grid of pole coordinates to randomly assign a pole orientation to each of our 10,000 synthetic objects. 

\begin{figure*}
\includegraphics[width=11.2cm, angle=0]{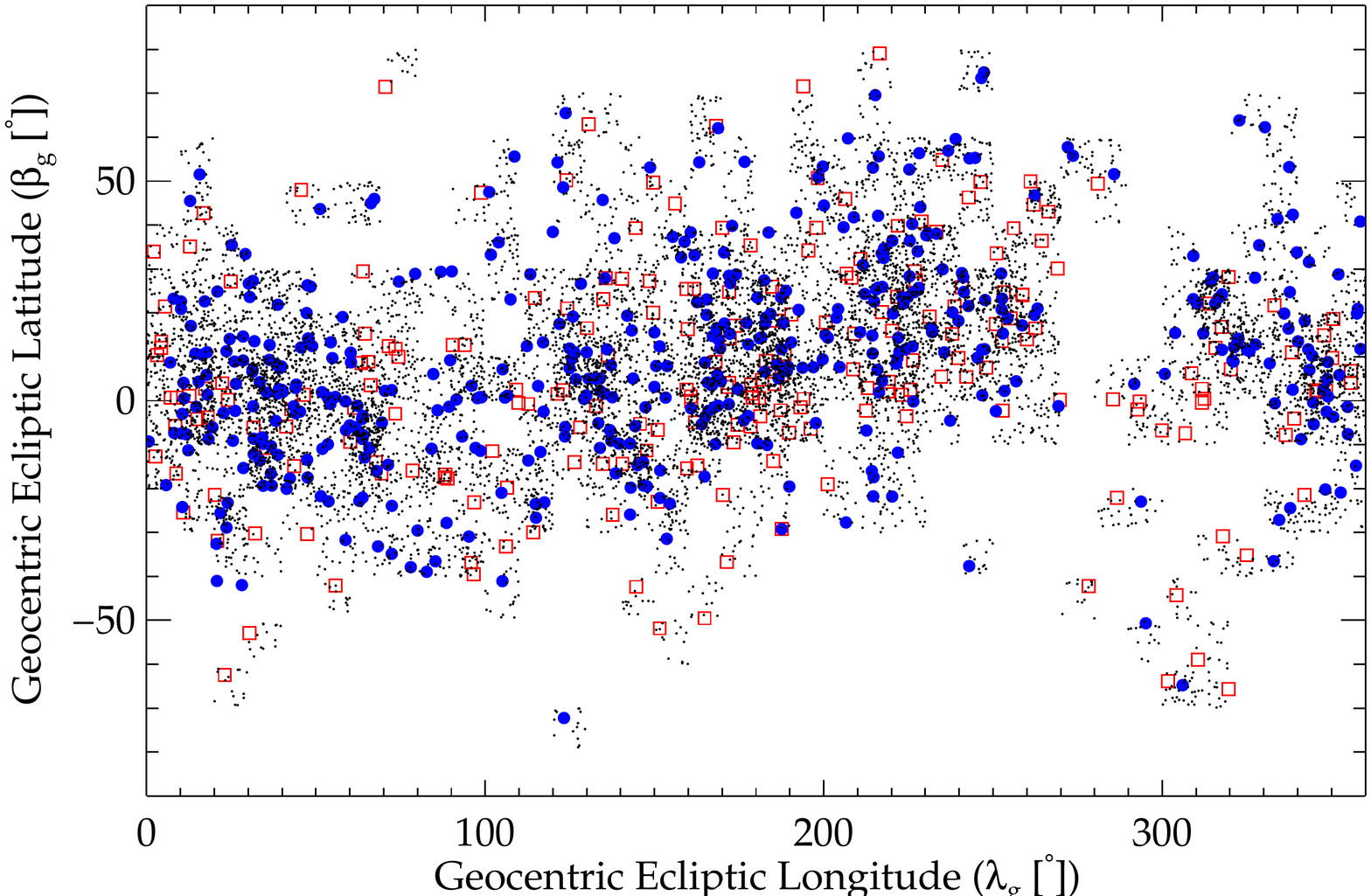}
\\
\includegraphics[width=11.2cm, angle=0]{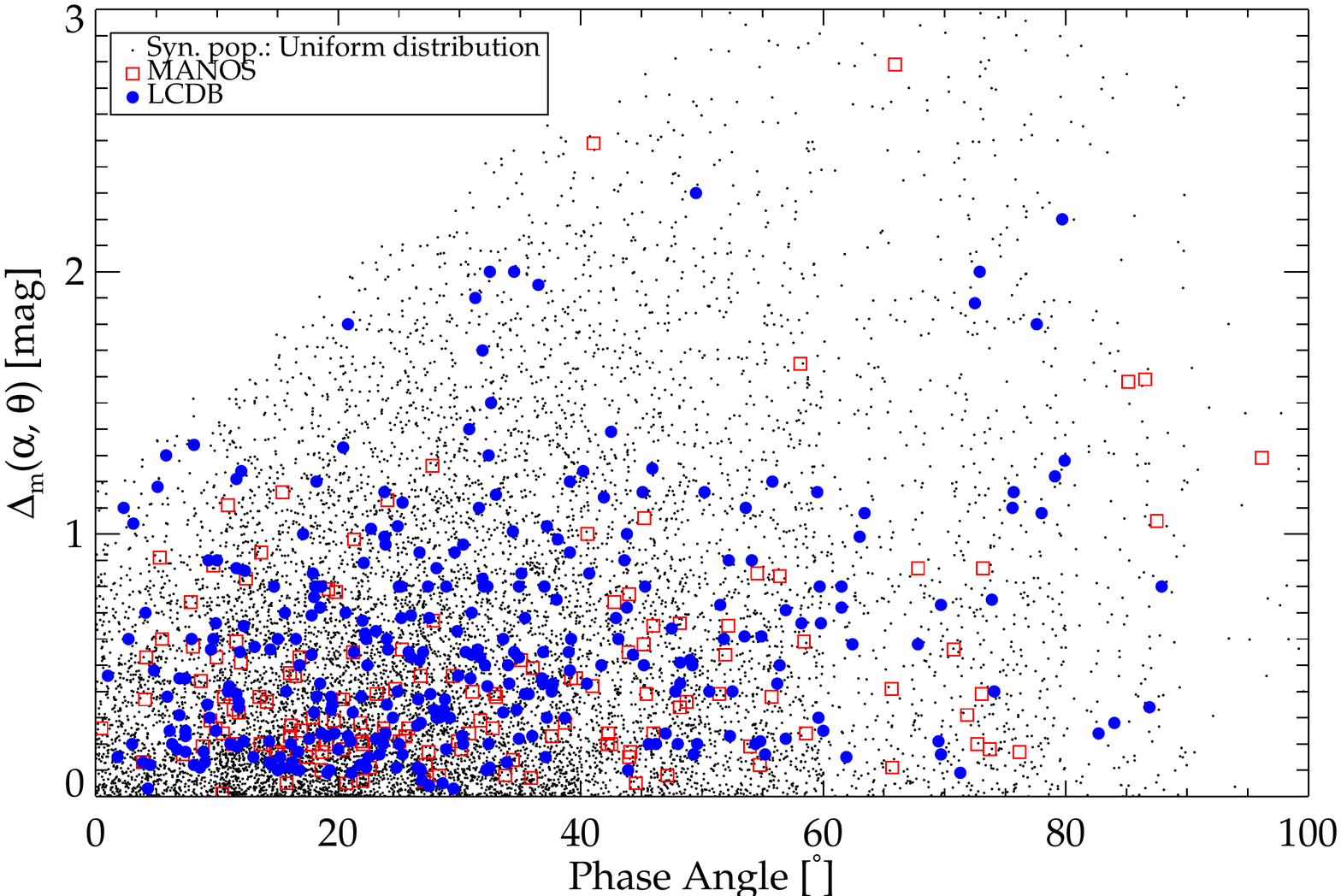}
\caption { We used the MANOS+LCDB sample (red squares+blue circles) to create a distribution of geocentric ecliptic coordinates for our synthetic population (black dots). As the aspect angle is unknown for our objects, we express the lightcurve amplitude as a function of phase angle. Following the procedure presented in Section~\ref{sec:synthetic}, the lower panel reports the lightcurve amplitude biased by phase angle of our synthetic population in the case of an uniform distribution of axis ratio. We overplotted the MANOS and LCDB samples for comparison. The synthetic population and the observations are in agreement.}
\label{Fig:uniform}
\end{figure*}

Equation~\ref{eq:theta2} also depends on the geocentric ecliptic coordinates ($\lambda_g$;$\beta_g$). For the MANOS sample, we use the zero phase of our lightcurve to estimate the ($\lambda_g$;$\beta_g$) of our objects. In order to present the most accurate sample, we also incorporated the LCDB objects with H$>$20~mag. Unfortunately, authors generally did not report the zero phase timing of their lightcurves. So, we used approximate coordinates for those objects based on the observing nights reported in the literature. Once ($\lambda_g$;$\beta_g$) are estimated for the MANOS+LCDB sample, we created a grid of geocentric ecliptic coordinates of 10$^\circ$$\times$10$^\circ$. Such a grid allowed us to take into account the approximate coordinates of the LCDB objects\footnote{In case of observations during close approach, some objects may move more than 10$^\circ$$\times$10$^\circ$ and thus are not in the right grid, however it is a small number of objects and this will not change the main conclusion of our simulations.}. Therefore, we created a distribution of ($\lambda_g$;$\beta_g$) based on the observations from the MANOS+LCDB sample (Figure~\ref{Fig:uniform}, upper plot). Using this and the distribution of ($\lambda_p$;$\beta_p$), we calculated the distribution of aspect angles (Equation~\ref{eq:theta2}) which is then used as input to Equation~\ref{eq:theta} to calculate a synthetic population of lightcurve amplitudes at zero phase. 

\paragraph{Step 2:}    

As the aspect angles of our observed sample are unknown, we cannot compare directly our dataset and the synthetic population. However, we can effectively observe our synthetic objects by assigning a phase angle based on the observed distribution of phase angles for MANOS+LCDB objects. By merging Equation~\ref{eq:alpha} and Equation~\ref{eq:theta}, we estimate the lightcurve amplitude of our synthetic population at these prescribed phase angles. In Figure~\ref{Fig:uniform}, we plot the MANOS sample and the LCDB objects with a H$>$20~mag and a phase angle lower than 100$^{\circ}$. We limit this analysis to small objects observed at a phase angle between 0 and 100$^\circ$ in order to mimic the MANOS sample. Based on Figure~\ref{Fig:uniform} (lower plot), it is obvious that the MANOS and LCDB observations are not uniform with phase angle. In fact, both data sets have an excess of objects observed at low/moderate phase angle (up to $\sim$40$^{\circ}$), and only an handful of objects are observed at high phase angle ($\alpha$$>$80$^{\circ}$). Drawing from the distribution of MANOS+LCDB objects, we create a non-uniform distribution of phase angles for our synthetic population (Figure~\ref{Fig:uniform}, lower panel), and then calculate the amplitude of our 10,000 synthetic objects.\\

In Figure~\ref{Fig:excess} (lower panel), we plot the normalized histogram of lightcurve amplitude for the synthetic population and the MANOS+LCDB samples. Error bars are $\sqrt{1/N}$ with N being the number of objects per bin. We limit our distribution to lightcurve amplitude up to 1.5~mag as only a handful of objects with higher lightcurve amplitude are reported. Generally, low lightcurve amplitude objects are difficult to obtain as they require a large amount of observing time under good weather conditions. In addition, observers have the tendency to not report or publish flat lightcurves. Therefore, there is a clear bias in the LCDB regarding these low amplitude objects, and thus we do not take into account objects with a lightcurve amplitude $<$0.1~mag. 

In Figure~\ref{Fig:excess} (lower panel), we plot our three synthetic populations (uniform distribution of b/a, an excess of spherical objects and an excess of elongated objects) for amplitudes between 0.1~mag and 1.5~mag. In order to compare the simulated population and the observed sample, we calculate the $\chi^2$ per degree of freedom: 
\begin{equation}
\frac{\chi^2}{\nu} = \frac{1}{\nu} \sum_{i}^{n} \Big[ \frac{f(\Delta m_i) - \Delta m_i}{\sigma_i} \Big]^2
\end{equation}
where $\nu$ is the degree of freedom, $\Delta m_{i}$ are the observed data, f($\Delta$ m$_{i}$) are the simulated results, and $\sigma_i$ are the uncertainties (i is the index of the bin and n is the bin number). Comparing the MANOS+LCDB data with the excess of elongated object distribution, we find a $\chi^2/\nu $ of 2.67. The MANOS+LCDB sample compared to the excess of spherical object distribution gives us a $\chi^2/\nu $ of 1.17, whereas compared to the uniform distribution the $\chi^2/\nu $ is 0.31. This suggests that a uniform distribution of b/a best fits the observed sample. Our model assumes a basic uniform distribution of b/a for prolate ellipsoids. Future improvements to this model could employ more realistic shapes based on radar observations and/or lightcurve inversion.

\section{Summary/Conclusions}

We report full lightcurves for 57$\%$ of our sample (82 NEOs), and constraints for the amplitude and period are reported for 21 NEOs. Thirty NEOs do not exhibit any periodic variability in their lightcurves. We also report 10 potential tumblers. 

MANOS found a potential new ultra-rapid rotator: 2016~MA. This object has a potential periodicity of 18.4~s. The confidence level of this periodicity is low and more data are required to confirm this result.  Unfortunately, there is no optical window to re-observe this object until 2025, and even then it only reaches V$\sim$22.5~mag. We also uncovered the fastest rotator to date, 2017~QG$_{18}$ rotating in 11.9~s.

Several MANOS targets display a flat lightcurve. Because of the well known relation between size and rotational period, we can infer that large objects (D$>$100~m) are slow rotators and their rotational periods were undetected during the amount of observing time dedicated. Based on this size dependent cut, we estimate that 43$\%$ of our flat lightcurves are slow rotators with a rotational period longer than our observing blocks. A flat lightcurve of a small NEO can be attributed to fast/ulra-rapid rotation which goes undetected because of the long exposing time used to retrieve a good signal-to-noise ratio. We suggest that 52$\%$ of our flat lightcures are potential fast/ultra-rapid rotators. We use the size of the object as a main criteria for these findings. This is an acceptable approximation, but may not be true for all the objects. 

We present a simple model to constrain the lightcurve amplitude distribution within the NEO population. One of the main parameters of our model is the b/a axis ratio of an object. We create several axis distributions, using an uniform distribution as well as an excess of spherical and elongated objects. Assuming that the pole orientation distribution reported in \citet{Vokrouhlicky2015} is representative of the NEO population, we generate 10,000 synthetic ellipsoids. We inferred that an uniform distribution of b/a best matches the observed sample. This suggests that the number of spherical NEOs is roughly equivalent to the number of highly elongated objects. 
 
A total of 78 MANOS objects are mission accessible according to NHATS which assumes a launch before 2040. However, considering only fully characterized objects, and NEOs rotating in more than 1~h, our sample of viable mission targets is reduced to 9 objects: 2002~DU$_{3}$, 2010~AF$_{30}$, 2013~NJ, 2013~YS$_{2}$, 2014~FA$_{7}$, 2014~FA$_{44}$, 2014~YD, 2015~FG$_{36}$, and 2015~OV. Two of these 9 objects will be bright enough during their next observing windows for new and complementary observations: 2013~YS$_{3}$ will have a V$\sim$18~mag in December-January 2020, and the visual magnitude of 2002~DU$_{3}$ will be 20.6~mag in November 2018.

\acknowledgments

Authors acknowledge the referee for useful comments to improve this work. 
Lowell operates the Discovery Channel Telescope (DCT) in partnership with Boston
University, the University of Maryland, the University of Toledo, Northern Arizona University and Yale University. Partial support of the DCT was provided by Discovery Communications. LMI was built by Lowell Observatory using funds from the National Science Foundation (AST-1005313).
This work is also based on observations obtained at the Southern Astrophysical Research (SOAR) telescope, which is a joint project of the Minist\'{e}rio da Ci\^{e}ncia, Tecnologia, e Inova\c{c}\~{a}o (MCTI) da Rep\'{u}blica Federativa do Brasil, the U.S. National Optical Astronomy Observatory (NOAO), the University of North Carolina at Chapel Hill (UNC), and Michigan State University (MSU). We also used the 1.3~m SMARTS telescope operated by the SMARTS Consortium. This work is based in part on observations at Kitt Peak National Observatory, National Optical Astronomy Observatory (NOAO Prop. ID:2013B-0270), which is operated by the Association of Universities for Research in Astronomy (AURA) under cooperative agreement with the National Science Foundation. This research has made use of data and/or services provided by the International Astronomical Union's Minor Planet Center. Authors acknowledge support from NASA NEOO grants NNX14AN82G, and NNX17AH06G. D. Polishook is grateful to the Ministry of Science, Technology and Space of the Israeli government for their Ramon fellowship for post-docs.

\clearpage


 
\clearpage

\startlongtable
\begin{longrotatetable}
\begin{deluxetable}{lccccccccc|ccc|ccccc}
\tabletypesize{\tiny}
\tablecaption{\label{Tab:Summary_photo} Observing log and results$^{a}$. }
\tablewidth{0pt}
\tablehead{
NEO  & UT-Date & $$Nb$_{im}$ & r$_{h}$& $\Delta$ & $\alpha$ & Filter &Tel & $\Delta T$& t$_{exp}$ & Rot. P. & $\Delta$m &  $\varphi_{0}$  & H &  D  & Dyn. &$\Delta$$_{V}^{SH}$& $\Delta$$_{V}^{NHATS}$\\
  &   &  &[AU]& [AU]& [$^\circ$]& && [min] & [s] & [h] &  [mag] & [2450000+] &   & [m] &   class & [km s$^{-1}$] & [km s$^{-1}$]\\
}
\startdata \textit{Full} & & & & & & &&& &		&			&	&    &   &	  & &  \\
  \textit{lightcurve} & & & & & & &&& &		&			&	&    &   &	  & &  \\
\textit{Symmetric} & & & & & & &&& &		&			&	&    &   &	  & &  \\
2014~UD$_{57}$ 	&  10/28/14  & 182  &  1.022   & 0.028-0.029   & 9.8-9.6   & wh  & KP4 & 153 &20 & 0.0959 &  0.88$\pm$0.02 	&  6958.62714  & 	 25.8  	& 20  &	Apollo    & 5.19 &  11.278  \\
2014~WF$_{201}$$^{b}$	& 12/01/14 & 44  & 1.010  &  0.028 & 29.4-29.5  & wh &KP4 & 28& 10&	0.4743 & 0.46$\pm$0.05 	&  6992.62728 & 	  25.6	&  22 &	Apollo    &  5.10& 7.094 \\
2017~LD & 06/04/17 & 53& 1.022&  0.0082 &15.9-16.1 & VR&DCT &21	& 3 & 0.0660	&		0.47$\pm$0.04	& 7908.83057	&   27.5 &  9 &	Amor  & 4.47 &8.339   \\
\textit{Asymmetric} & & & & & & & &		&			&	&    &   &	  & &  \\
1999~SH$_{10}$	& 03/28/14 &  178 &  1.120  & 0.147-0.146  &  31.7 & wh & KP4 &201 &35 &	0.1264 &  0.29$\pm$0.03 	&  6744.87878 & 	 22.6	& 89  &	Apollo   & 5.57 &8.634 \\
2006~HX$_{30}$	&  05/27/15  & 73  &  1.049-1.050 &  0.040 & 24.7  & r' & SOAR &93 &20&0.0966 &  0.41$\pm$0.03  	& 7169.74255  & 	26.2 	& 17  & Amor  & 4.72 &10.456 \\
2010~MR	& 07/11-14-21/14  & 126  &  1.494-1.432   &  0.478-0.418  &  2.9-4.9  & V & CTIO &67;252;120&60& 2.42 &   0.13$\pm$0.05 	&  6849.80986 & 	 18.6	& 566  &	Amor   & 6.80 & -  \\
2012~BF$_{86}$	& 02/22/16 &  91 & 1.043  & 0.0825  & 46.7-49.6  & VR & DCT &71&15&	0.0491 & 0.34$\pm$0.04 	& 7440.68168  & 	22.6 	& 89  &	    Aten &  10.14 & -\\
2013~SB$_{21}$	& 10/14/13 &  64 &  1.031 & 0.034  &  12.4-12.5 & wh &KP4&49&35 &	0.0584 & 0.83$\pm$0.04 	&  6579.77774 & 	  27.0	& 11  &	 Amor   &  4.47 & 8.588 \\
2013~SR	& 10/14/13 &  51 & 1.050  & 0.070  & 40.5-40.6  & wh &KP4 &47&10&	0.1305 & 1.00$\pm$0.03 	& 6579.63927  & 	  24.1	&  44 &	 Amor   &  5.27& - \\
2013~TL	& 10/14/13 &  89 & 1.022  & 0.085  &  70.7-70.8 & wh &KP4 &57&5&	0.7942 & 0.56$\pm$0.04 	& 6579.97030  & 	  22.2	&  107 &	 Apollo   & 5.89& - \\
2014~FF	& 03/28/14 & 161  & 1.030  & 0.040  &  35.9-36.2 & wh & KP4 &72&8&	0.1032 & 0.49$\pm$0.03  	&  6744.68831 & 	  24.2	& 42  &	Amor    & 6.51 & -\\
2014~FR$_{52}$	&  04/18/14  & 85  & 1.135  &  0.148-0.147 & 20.1  & wh & KP4 &46&15&  0.0149 & 0.37$\pm$0.06    	&  6765.66520 & 	  23.9	& 49  &	Amor    &  6.09& -\\
2014~HB$_{177}$	& 05/06/14  & 21  &  1.009 & 0.0034  & 87.1-87.8  & VR  & DCT &11&10&	0.0239  & 1.05$\pm$0.06  	& 6783.97448  & 	  28.1 	&  7  & Apollo  & 4.83 & 6.752 \\
2014~HE$_{177}$	&  05/06/14  &  69 & 1.056  & 0.049  &  17.0-16.9 & VR & DCT &55&15& 0.0897 &  0.25$\pm$0.03  	& 6783.65329  & 	25.8	& 20  &	    Amor & 5.89 & -\\
2014~HF$_{5}$	&  05/06/14  & 37  & 1.055  &  0.056 & 33.7-33.8  & VR & DCT &45&20& 0.1038 &  0.08$\pm$0.03  	&  6783.91201 & 	25.3 	& 25  &	    Amor& 5.92& - \\
2014~HN$_{178}$	&  06/16/14  & 144  & 1.032  &  0.044 & 67.7-67.9  & r' & SOAR &142&10& 0.0367 &  0.87$\pm$0.02  	& 6824.72654  & 	23.5 	& 59  &	    Amor &  5.58& -\\
2014~JD	&  05/06/14 & 76  &  1.037 &  0.030 &  20.3-20.5 & VR & DCT &52&7& 0.0714 &  0.18$\pm$0.04  	& 6783.74905  & 	26.3 	& 16  &	  Apollo & 6.13 & -\\
2014~JJ$_{55}$	&  06/03/14  & 52  & 1.073  & 0.086  & 45.2-45.3  & VR & DCT &168&120& 0.915 &  1.06$\pm$0.03  	&  6811.71253 & 	25.3 	&  25 &	    Apollo & 4.94 & 6.380\\
2014~JR$_{25}$	&  05/10-11/14  & 53  &  1.048-1.057 & 0.043-0.052  &  26.2-25.3 & V & CTIO &106;75&45& 0.487 &  0.26$\pm$0.04  	& 6787.73564  & 	23.4 	&  62 &	    Apollo &  7.65 & - \\
2014~KA$_{91}$	& 06/04/14 & 64  & 1.034  & 0.026  &  40.9-41.0 & wh & KP4 &32&6&	0.120 & 0.42$\pm$0.04 	& 6812.65859  & 	  25.5	&  23 &	 Apollo   & 5.57 & 11.846\\
2014~KH$_{39}$	&  06/03/14  & 55  & 1.016  &  0.0051-0.0050 &  65.3-66.5 & VR & DCT&50&1 & 0.0440 &  2.79$\pm$0.02  	& 6811.66803  & 	26.2 	&  17 &	    Apollo & 6.53 & - \\
2014~OV$_{3}$	& 02/10/15 &  56 & 1.155  & 0.171  & 9.9-10.0  & wh & KP4 &121&70&	0.3491 & 0.53$\pm$0.02 	& 7063.73463  & 	23.2	& 68  &	  Apollo & 4.73 & 7.046\\
2014~TM$_{34}$	&  10/17/14  &  106 & 1.047  & 0.053-0.054  & 18.6-18.7  & VR & DCT &57&15& 0.0249 &  0.17$\pm$0.03  	&  6947.88390 & 	25.0	& 29  &	    Amor & 5.58 & -\\
2014~TP$_{57}$	&  10/17/14  & 71  &  1.024 & 0.028  &  15.6-15.5 & VR & DCT &29&8& 0.0137 &  0.16$\pm$0.04  	&  6947.64265 & 	26.4 	& 15  &	 Amor    &  5.18& - \\
2014~UX$_{7}$ 	&  10/28/14  &  85 &   1.068  &  0.075  &   10.5-10.6 & wh  & KP4 &102&45&  0.0366 &    0.38$\pm$0.04	&  6958.84348  & 	25.6   	& 22  &	 Amor   & 5.26 &  7.294  \\
2014~WU$_{200}$$^{c}$	& 12/01/14 &  60 & 0.996  &  0.011 & 16.1  & wh & KP4 &41&10&	0.0179 &  0.27$\pm$0.05 	& 6992.72115  & 	 29.1 	& 4  &	Apollo    &  4.17& 4.206\\
2014~YT$_{34}$	&  01/13/15  & 124  & 1.012  & 0.037  &  38.9-39.5 & r' & SOAR &133&5& 0.1806 &  0.45$\pm$0.02 	&  7035.55787 & 	24.7 	& 34  &	 Apollo    & 5.75 & 9.089\\
2015~CF	&  02/11/15  & 157  & 1.069  & 0.086  & 15.7-15.8  & VR & DCT &56&7& 0.1841 &  0.05$\pm$0.03  	&  7064.87459 & 	23.5 	& 59  &	 Amor    &  5.99 & -\\
2015~HS$_{11}$	& 04/25/15 & 44  &  1.029 & 0.023  &  4.0-4.1 & wh & KP4 &44&7&	0.0193 & 0.37$\pm$0.04 	&  7137.73600 & 	27.1 	&  11 & Amor &	 4.29& 6.683 \\
2015~HU$_{9}$	& 05/08/15  & 128  & 1.099-1.098  & 0.130-0.129  &  43.8-44.2 & wh & KP4 &145&30&	0.2130  &  0.15$\pm$0.04  	& 7150.66205  &  23.4 	&  62  & Apollo &8.24 & -\\
2015~HV$_{11}$	& 05/08/15 & 55  &  1.072   &  0.068  &  22.1-22.0  & wh & KP4 & 61&25&	0.3102 & 0.20$\pm$0.04 	&  7150.89848 & 	24.1 	&  44 & Amor &	6.12  & - \\
2015~JD 	& 05/08/15 &  98 &  1.022 &  0.016 &  36.1-35.7 & wh &KP4 &72&5&	0.0339 & 0.07$\pm$0.03  	&  7150.79977 & 	25.5  	& 23  &	Apollo    &  6.00 & 11.965\\
2015~KE 	& 05/22/15 & 100  & 1.043  & 0.034  &  27.4-27.5 & VR & DCT &66	&5&0.0562 & 0.17$\pm$0.03 	&  7164.65626 & 	26.4 	&  15 &	Aten     &  4.54 & 4.296\\
2015~KM$_{120}$	&  05/26/15  & 166  & 1.042  & 0.050  &  54-53.9 & wh & KP4 &110&15& 0.0296  &   0.19$\pm$0.04   	&  7168.66447 & 	24.7 	&   34  &	Amor     &  6.62 & -\\
2015~KO$_{122}$	&  05/27/15  & 43  & 1.036-1.037  &  0.024 &  14.9-14.8 & r' & SOAR &39&10& 0.0648 &  0.17$\pm$0.04  	& 7169.83972  & 	27.0 	& 11  &	Apollo    &  6.78 & -\\
2015~KQ$_{120}$	&  05/27/15  &  62 & 1.040  &  0.029 &  21.2-21.4 & r' & SOAR &58&10& 0.0898 &  0.98$\pm$0.04  	&  7169.87814 & 	26.7 	&  13 &	Apollo    &  4.89 & 10.883\\
2015~KW$_{120}$	&  05/26/15  &  159 &  1.037 & 0.024  &  10.5 & wh & KP4 &63&3&  0.1355 &   0.24$\pm$0.03   	&  7168.76662 & 	26.0 	& 18   &	 Apollo    &  6.20 & -\\
2015~MX$_{103}$	&  06/29/15  & 129  & 1.045  &  0.039 &  42.2 & r' & SOAR &111&8& 0.7865 &  0.20$\pm$0.04  	& 7202.58034  & 	24.4 	& 39  &Amor	    & 5.36 & - \\
2015~RF$_{36}$	& 09/14/15 &  146 & 1.053-1.052  &  0.057 &  34.5-34.3 & wh & KP4 &100&15&	 0.0123 &  0.14$\pm$0.02  	&  7279.93330 & 	23.4 	&  62  &	   Aten   & 5.63 & 6.312\\
2015~RQ$_{36}$	& 09/14/15 &  69 & 1.058  & 0.058  &  25.6-25.5 & wh & KP4 &59&20&	 0.4812 &  0.23$\pm$0.04  	&  7279.87219 & 	24.5 	&  37  &	   Apollo   & 5.13& - \\
2015~TL$_{238}$	& 10/24/15 & 52  & 1.035  & 0.042  & 16.4-16.5  & VR & DCT & 47	&5&0.0713 & 0.46$\pm$0.04 	&  7319.68601 & 	24.9 	& 31  &	Apollo     & 7.43& - \\
2015~VE$_{66}$	&  11/24/15  & 126  & 1.005  & 0.021  &  32.7-33.3 & r' & SOAR&101&2 & 0.037 &  0.38$\pm$0.02  	& 7350.71888  & 	24.1 	& 44  &	    Amor &  6.03& - \\
2015~XF	&  12/29/15  &  98  & 1.059  & 0.078  &  11.9-12.0 & r' & SOAR &88&20& 0.1003 &  0.51$\pm$0.03  	&  7385.75211 & 	24.4 	& 39  &	 Amor    & 6.90& - \\
2016~AD$_{166}$	& 01/19/16 & 85  & 1.074  &  0.101 & 25.1-25.0  & VR & DCT &49&15&	0.0085 & 0.21$\pm$0.03 	&  7406.76968 & 	23.6 	& 56  &	Apollo     & 7.13& - \\
2016~AF$_{166}$	& 01/19/16 &  37 & 0.995  &  0.027 &  65.6-65.7 & VR & DCT &14	&10&0.0278 & 0.41$\pm$0.05 	& 7406.80902  & 	25.4 	&  24 &	  Apollo   & 6.30 & -\\
2016~AO$_{131}$	& 01/19/16 &  65 & 1.054  & 0.093  & 39.6  & VR & DCT &27&10&	0.0216 & 0.45$\pm$0.04 	&  7406.99723 & 	24.1 	&   44 &Apollo 	    & 4.57 & 9.651\\
2016~AU$_{9}$	& 01/12/16 &  59 &  1.061 &  0.079 &  11.6 & wh & KP4 &135	&30&0.594 & 0.59$\pm$0.04 	& 7399.87573  & 	25.4 	& 24  &	 Amor   & 6.03 & -\\
2016~AV$_{164}$	& 01/19/16 &  40 &  1.059 &  0.078 & 13.6  & VR & DCT &15&10&	0.0123 & 0.20$\pm$0.04 	&  7406.94935 & 	24.9 	& 31  &	Amor    &  6.24& -\\
2016~CS$_{247}$	& 02/22/16 & 108  & 1.019  &  0.033-0.034 &  27.2 & VR & DCT &42&7&	0.0514 & 0.12$\pm$0.03 	& 7440.85852  & 	25.6 	& 22  &	  Apollo   & 4.43 & 8.076\\
2016~EL$_{157}$	&  03/16/16  & 50  &  1.013 &  0.018 &  8.7-8.9 & VR & DCT &20&1& 0.0198 &  0.19$\pm$0.03  	&  7463.89240 & 	27.1 	&  11 &Apollo	    & 6.36 & -\\
2016~EN$_{156}$	&  03/16/16  & 94  &  1.009 & 0.014  &  8.7-8.6 & VR & DCT &27&2& 0.0863 &  0.44$\pm$0.05  	& 7463.91622  & 	27.8 	&  8 &	Apollo     & 4.98 & 9.108\\
2016~FL$_{12}$	& 04/07/16 & 46  & 1.032  &  0.032 &  17.3-17.4 & r' &SOAR &47&9&	0.3333 & 0.15$\pm$0.05 	& 7485.71234  & 	  26.3	& 16  &	 Apollo   &  4.74 & 8.672\\
2016~FZ$_{2}$	&  03/22/16  &  139 &  1.1053 & 0.062  &  23.8-23.7 & VR & DCT &49&5& 0.0387 &  0.26$\pm$0.05  	&  7469.96281 & 	24.5 	&  37 &	Amor    & 6.62 & -\\
2016~GW$_{221}$	& 04/19/16 &  113 &  1.041 &  0.043 &  30.3-30.0 & r' & SOAR &98&3&	0.2856 & 0.18$\pm$0.05 	&  7497.85083 & 	  24.8	&  32 &	Aten    & 7.43 & 9.526\\
2016~JP$_{17}$	& 05/09/16 & 95  & 1.029  & 0.032  & 52.3-52.0  & VR &DCT &27&1&	0.0702 & 0.65$\pm$0.02  	& 7517.78384  & 	  23.1	&  71 &	Apollo    & 6.17 & -\\
2016~MA 	&  06/17/16  & 51  &   1.023  &  0.012  &  54.7-54.9  & VR  & DCT &29&5&  0.0051 &    0.12$\pm$0.04	& 7556.74356   & 	 27.5  	& 9  &	 Apollo   & 5.39 &  11.229  \\
2016~NG$_{38}$ & 07/18/16 & 101 & 1.038 & 0.034 & 48-48.3 & r' & SOAR &114&7&	2.47	&	0.66$\pm$0.04	& 7587.85278	&   25.1 &  28 &	Amor  & 5.98&  -  \\
2016~NK$_{39}$ & 08/15/16 & 74 & 1.070 & 0.085& 45.9-46.1 &r' & SOAR &	107&30&1.46	&	0.24$\pm$0.05		&7616.49071 &   23.9 & 49  &	Amor  & 5.77& 11.025 \\
2016~PA$_{40}$ & 08/16/16 & 98 & 1.092 & 0.082 & 13.7-13.6   & r'& SOAR&	77&10&	0.1375 &		0.93$\pm$0.03	& 7616.86135	&   24.4 &  39 &	Apollo  & 7.16& 11.513 \\
2016~PP$_{27}$ & 08/16/16 &116 & 1.080 & 0.081-0.082 & 32.7-32.8 &r' & SOAR&95&5&	1.55	&		0.26$\pm$0.07	&7616.69028	&   23.6 &  56 &	Apollo  & 7.12& - \\
2016~RB$_{1}$ & 09/07/16 & 198 & 1.009  & 0.0016-0.0012 & 30.9-28.5 & VR &DCT &153&3&	0.0267	&		0.21$\pm$0.03	& 7638.82497	&   28.3 &  6 &	Aten  & 7.49& 8.626 \\
2017~EA$_{3}$ & 03/09/17& 130 &1.042 &0.071  &44.6-44.5 & VR & DCT&65&5&	0.71	&	0.05$\pm$0.03		&7821.75821	&   23.2 &  68 &	Apollo  & 6.72& - \\
2017~EE$_{4}$ & 03/15/17 &106 & 1.000 &0.019 & 71.6-72.1 &VR & DCT&30&1&0.00699	&	0.31$\pm$0.05		&7827.99828	&   25.0 &  29 &	Apollo  & 6.00&  -\\
2017~EH$_{4}$ & 03/09/17 & 139& 1.056& 0.070& 25.3-25.2 & VR &DCT &55&5&0.0624		&	0.56$\pm$0.02		& 7821.69563	&   24.1 &  44 &	Amor  & 5.67& - \\
2017~FJ      &  03/19/17 &179 &1.009-1.008 & 0.013& 8.0& VR & DCT&	49&5&0.0165&	0.57$\pm$0.03		&7831.88047	&   28.2 &  6 &	Apollo  & 5.51& 11.780 \\
2017~FK      & 03/19/17 & 108 & 1.005 &0.085-0.084 & 17.7-18.1& VR& DCT&	27&2&0.00428	&	0.30$\pm$0.03		&7831.72251	&   27.3 &  10 &	Apollo  & 5.61& 9.042 \\
2017~QG$_{18}$       & 08/27/17 &296 & 1.032 & 0.024-0.023 & 22.0-21.8 & VR&DCT &91&3&	0.003298	&	0.21$\pm$0.05		&7992.64725	&   27.0 & 11  &	Apollo  &5.26 &  11.411\\ 
2017~QK  & 08/21/17 &47 &1.151 & 0.147& 17.9 & VR&SOAR &57&30&	0.1599	&	0.20$\pm$0.07	&7986.54836&  23.8 &  51 &	Apollo  & 5.45& - \\
2017~QT$_{1}$       &08/21/17 & 158 &1.030 &0.020-0.019 & 19.5-20.1 & VR & SOAR&70&10&	0.77	&0.78$\pm$0.05	&7986.69346	&   26.7 &  13 &	Apollo  & 8.07& - \\  
\textit{Complex} & & & & & & & &		&			&	&    &   &	  & &  \\
2014~HS$_{184}$	& 06/02/14   & 89  & 1.069  &  0.061-0.060 &  24.0-24.1 & V & CTIO &233&20& 2.02 &  1.13$\pm$0.04  	&  6810.56578 & 	23.3 	& 65  &	    Amor &  5.72 & -\\
2014~HW 	&  04/24/14  &  126  & 1.014   & 0.0089-0.0087  &  15.4  & VR & DCT &108 &4&0.0641 &  1.16$\pm$0.03   	& 6771.81914  & 	 28.4 &  6 &Apollo &  4.54 & 5.690 \\
2016~BF$_{1}$	& 01/19/16  & 75  & 1.010  &  0.035 & 41.1-41.0  & VR  & DCT &54&10&	0.2624  &   2.49$\pm$0.02	&  7407.02439 & 	  25.4	& 24  & Apollo  & 5.59 & - \\
2016~DK	& 02/22/16 & 285  &  1.048 & 0.108  & 54.5-54.6  & VR & DCT &131&10&	1.30 &  0.85$\pm$0.02 	&  7440.94528 & 	22.4 	& 98  &	  Amor  & 11.34 & -\\
2016~ES$_{1}$	& 03/16/16   &  99 &  1.089 &  0.095 & 9.9  & VR & DCT & 38&3&0.3484 &  0.26$\pm$0.04  	& 7463.94721  & 	24.1 	&  44 &	 Amor   &  6.65& -\\
2017~EK  & 03/15/17 &35 &0.995 &0.017 &86.5-86.6 & VR& DCT&10&1&	0.0064&	1.59$\pm$0.05	&	7827.99072 &   24.1 &44   &	Apollo  & 7.03&  -\\
2017~EZ$_{2}$ &03/09/17 & 118 & 1.008& 0.015& 15.8-16.4 & VR &DCT &39&3&0.0138	&	0.23$\pm$0.03	& 7821.80369 &   25.1 & 28  &	Amor  &8.75 &  -\\
2017~HV$_{3}$ & 08/27/17 &278  &1.081 & 0.102 & 44.1-43.9 &VR  &DCT &163&7&0.0881		&	0.77$\pm$0.03		&7992.87174	&   23.7 &  54 &	Amor  & 4.65& 8.359 \\
2017~JM$_{2}$ & 05/14/17 &225 & 1.021&  0.015 &45.7-46.3& VR& DCT&67&2&0.0188		&	0.65$\pm$0.02		&7887.81127	&   24.3 &  41 &	Apollo  & 9.05&  - \\
2017~KJ$_{27}$ & 05/28/17& 76& 1.023&0.019 & 60.4-61.0& VR&DCT &19&2&0.036		&0.58$\pm$0.03		&	7901.94000&   25.4 & 24  &	Apollo  & 7.16& - \\
2017~LE       & 06/04/17 & 224 & 1.031 & 0.019 & 33.1-32.7  &VR & DCT&77&2&	0.0281	&	0.39$\pm$0.06		&7908.85535	&   26.4 &  15 &	Amor  & 6.22& - \\
2017~MO$_{8}$ & 07/03/17 &108 &1.018-1.017 & 0.011& 84.8-85.5 & VR& DCT&34&1&0.0544		&1.58$\pm$0.03			&7937.75364	&   26.0 &  18 &	Apollo  & 6.52&-  \\
2017~QX$_{1}$       & 08/21/17 &283 &1.050 & 0.039 & 10.8-11.0 & VR & SOAR&85&5&	1.34	&	1.11$\pm$0.02		&7986.86290	&   24.8 &  32 &	Amor  &5.53 & - \\ 
\hline
\hline
\textit{Partial} &		& & & & & & &	&		&&	&	&		 	&  & \\
\textit{lightcurve} &		& & & & & & &	&	&&		&	&		 	&  & \\
2013~VY$_{13}$ & 01/03/14   &36   & 1.458  & 0.519  &  19.4 &  wh & KP4 &118&180& $>$2  & $>$0.1   	&  6660.72197 & 	  21.2	&  171 &	Apollo    &  6.30 & -\\
2013~XX$_{8}$	& 02/05/14 & 133 & 1.072  & 0.087  &  9.2-9.3 & wh & KP4&128	&30-45&$>$2.5 &  $>$0.6 	& 6693.72231  & 	  24.4	& 39  &	Amor    & 4.57& 10.364 \\
2013~YS$_{2}$	& 01/27/14 & 122  & 1.023  &  0.057 &  45.9-45.8 & wh & KP4 &134&15&	$>$2  & $>$1.0  	&  6684.60839 & 	  23.3	& 65  &	Amor    &  4.77& 10.346\\
2014~FA$_{7}$	& 03/28/14 &  73 &  1.025-1.026 & 0.030  & 24.3  & wh & KP4 &	127&60&$>$2 &  $>$0.2 	&  6744.78622 & 	  26.7	&  13 &	Apollo    & 5.17& 7.232 \\
2014~HK$_{129}$	& 05/08/14 &  164 &  1.198-1.199 & 0.261  &  39.1 & wh &KP4 &95&35&	$>$3 &  $>$0.6 	& 6785.63865  & 	  21.1	&  179 &	Apollo    & 6.26 & - \\
2014~WA$_{366}$	& 12/27/14 &  21 &  1.042 &  0.060 &  13.1 & wh &KP4 &19&45&	$>$0.5 &   $>$0.5	& 7018.93337  & 	  26.9	& 12  &	Apollo    & 4.14 & 4.568\\
2014~WO$_{69}$	& 12/01/14 &  66 &  1.131-1.131 & 0.149  &  11.2-11.3 & wh &KP4 &164&90&	$>$2.5 &  $>$0.2 	&  6992.83219 & 	  23.6	&  56 &	Amor    & 6.19 & -\\
2015~AA$_{44}$	&  02/10/15  & 116  & 1.010  & 0.052  & 61.8-62.1  & wh & KP4 &71&7&  $>$1 & $>$0.4   	& 7063.66223  & 	23.9 	&  49 &	  Apollo   & 5.68& - \\
2015~ET 	&  03/14/15  & 132  & 1.018  &  0.026 & 20.0-19.9  & wh & KP4 & 96&30&$>$2 &  $>$0.5   	&  7095.88517 & 	 26.7 	&  13   &   Apollo  & 6.48 & -\\
2015~GC$_{14}$	& 04/25/15 & 17  &  1.095 &  0.095 & 19.6  & wh & KP4 &39&40&	$>$0.5 &  $>$0.3 	&  7137.66520 & 	  24.8	&  32 &	 Amor   & 5.29 & -\\
2015~PT$_{227}$	&  08/30/15  & 65  & 1.018-1.017  &  0.025 & 71.2-71.5  & r' & SOAR &40&3& $>$1  &  $>$2.0   	& 7264.88016  & 	23.9 	&  49  &Apollo    &  6.29& -\\
2015~QA 	&  09/03/15  & 85  &  1.112 &  0.110 &  19.0 & VR & DCT &64&10&  $>$1  &   $>$0.2   	& 7268.68778  & 	 22.9 	& 78  &	   Amor &  6.56 & -\\
2015~XA$_{379}$	&  01/12/16  &  9 & 1.026  &0.049   & 27.1  & wh & KP4 &11&40&  $>$0.2 &   $>$0.1  	& 7399.77151  & 	 25.4 	& 24  &	   Amor &  4.22 & 7.629\\
2016~GF$_{216}$	& 05/17/16 &  19 &  1.058 &  0.051 & 24.7-24.6  & VR & DCT &17&20&	$>$0.5 &  $>$0.2 	&  7525.85932 & 	  24.9	& 31  &	Amor    & 4.62& 9.009 \\
2016~HN$_{2}$	& 05/09/16 &  95 & 1.103  & 0.114  &  33.3 & VR &DCT &54&10-15&	 $>$1 &  $>$0.2 	&  7517.92273 & 	  23.5	&  59 &	Apollo    &  6.08& -\\
2016~HP$_{3}$	& 05/22/16 & 169  & 1.041  & 0.050  & 54.0-54.2  & VR &DCT & 60 &5&	$>$1  &  $>$1.0 	& 7530.83529  & 	  23.7	& 54  &	Amor    & 6.46 & -\\
2016~JD$_{18}$	& 05/09/16 &  93 &  1.069 &  0.066 &  23.8 & VR &DCT &36&10&$>$0.5	 & $>$1.3  	&  7517.85625 & 	  24.7	&  34 &	Apollo    & 7.30 & -\\
2016~JE$_{18}$	& 05/09/16 &  176 &  1.036 & 0.027  &  9.4-9.3 & VR &DCT &58&3-5&	$>$1 &  $>$0.6 	&  7517.80558 & 	  26.3	& 16  &	Amor    & 5.88& - \\
2016~LO$_{48}$	& 06/15/16 & 72  & 1.041  &  0.032 & 36.5-36.6  & VR &DCT & 28 &5-6&  $>$0.5 & $>$0.4  	& 7554.78964  & 	  25.4	& 24  &	Amor    & 5.44 & 9.817\\
2017~EK$_{3}$ & 03/09/17 &216 &1.025 & 0.033& 14.1-13.8 & VR &DCT &82&7-10&	$>$1.5	&	$>$0.5	& 7821.85281&   26.3 & 16  &	Apollo  & 5.87& 8.840 \\
2017~QU$_{17}$  &08/27/17  & 317&1.059 & 0.052& 20.8-20.9& VR & DCT&105&5&	$>$2	&	$>$0.2		& 7992.79683	&   26.1 &17   &	Amor  &6.20 & - \\ 
\hline\hline
\textit{Flat} &		& & & & & & &	&			&	 	&&	&&&	\\
\textit{lightcurve} &		& & & & & & &	&			&	 	&&	&&&	\\
2008~HB$_{38}$	&  10/28/13  & 92  & 1.276-1.277  & 0.297-0.298  & 15.7  &  r'  & KP2 &136&40-60&  - & -   	& 6593.75162  & 	  21.1 	&  179 &	Apollo    & 5.73& - \\
2010~CF$_{19}$	& 08/16/13 & 11  &  1.127 & 0.128  &  25.2 & V & CTIO &20&30&	-& -  	&  6520.71312 & 	  21.7	&   135 &	Apollo  & 5.46 & 9.449 \\
2012~ER$_{14}$	& 02/05/14 &  56 & 1.348  & 0.404-0.405  & 22.4  & wh &KP4 &145&90&	-&   -	&  6693.60084 & 	 20.5 	& 236   &	Amor    & 5.43& -  \\
2013~PR$_{43}$	& 09/17/13 & 24  &  1.159-1.160 &  0.158 &  11.7-11.5 & r' &SOAR &62&120&	-& -  	& 6552.76646  & 	 23.4 	&  62  &	Apollo  &  5.11 & -\\
2013~SY$_{19}$	& 10/10/13 &  61 & 1.128  & 0.130  & 4.0  & r' &SOAR &116&60&	-& -  	&  6575.76236 & 	 24.8 	&   32 &	Amor  &  4.95 & -\\
2013~UE$_{1}$	& 10/30/13  & 97  & 1.033  & 0.049  &  35.0-35.2 &  r' & KP2 &147&20-30&-	& -  	&  6595.72699 & 	 24.4 	& 39   &	Apollo  & 5.96 & - \\
2013~UE$_{3}$	& 10/30/13 &  49 &  1.137 &  0.145 &  6.1-6.0 & r'  &KP2 &97&50&	-&  - 	&  6595.86569 & 	  22.7	&  85  &	 Apollo &   5.60&  7.438\\
2013~XY$_{20}$	& 01/03/14 &  118 &  1.011 &  0.045 &  52.1 & wh & KP4&127&30 &	-&-   	&  6660.97473 & 	  25.5	&  23  &	 Amor & 3.98  & 6.507\\
2014~CS$_{13}$	&  03/25/14  & 30  & 1.235  & 0.257-0.258  &  20.1-20.2 & VR & DCT &159&300& -  &-    	&  6741.77483 & 	24.0 	& 47   &Apollo    & 5.36& 8.754 \\
2014~KL$_{22}$	&  06/03/14  & 67  &  1.053 & 0.047  &  34.2 &  VR  & DCT  &41&10&  - & -   	& 6811.86838  & 24.6	   	&  35 &	Amor    &  5.76 & 11.740 \\
2014~OA$_{2}$	& 08/01/14  &  104 & 1.176  &  0.162 & 6.7-6.8  & V & CTIO &145&30&	-& -  	&  6870.72426 & 	  21.3	&   163 &	Amor  & 5.86 & - \\
2014~QV$_{295}$	&  09/16/14  &  96  & 1.074  & 0.072  & 15.8  & VR & DCT &76&10&  - & -   	&  6916.94185 & 	24.9 	& 31  &	  Amor  & 6.32 & -\\
2014~TR$_{57}$	&  10/17/14  &  152 & 1.030  & 0.037  & 26.3-26.4  & VR & DCT &95&6-15&  -  & -    	&  6947.78720 & 	25.2	&  27 &	  Amor  &  5.48& - \\
2014~UY$_{7}$ 	&  10/28/14 & 62  & 1.045    & 0.065   &  36.4  & wh  & KP4&98&35-40 &  - &    -	&  6958.93640  & 	   24.8  &	 32   & Amor & 6.44  &  - \\
2014~WU$_{200}$$^{b}$	&  11/26/14  &  68 & 1.003  & 0.016  & 9.1  & VR & DCT &93&45-55&  -&   -  	& 6987.79201  & 	29.1 	& 4  &	 Apollo    & 4.17& 4.206 \\
2014~WX$_{202}$	&   11/27/14 &  179 & 0.994  & 0.0075   & 13.4-13.3  & VR & DCT &68&5&  -&-     	& 6988.95925  & 	29.6 	&  3 &	Apollo     &  4.09& 4.151 \\
2015~KT$_{56}$	& 05/26/15   & 91  &  1.063 & 0.050-0.051  & 8.4-8.5  & wh & KP4 &73&15-20& - &  -  	& 7168.81786  & 	26.1 	&  17 &	 Apollo    & 6.25 & -\\
2015~KV$_{18}$	& 05/26/15   &  108 &  1.135 &  0.127 &  14.3 & wh & KP4 &98&25& - &  -  	& 7168.87484  & 	23.8 	& 51 &	 Amor    & 5.97& - \\
2015~LK$_{24}$	&  06/29/15  &  201 & 1.040  & 0.060  &  65.0-65.3 & r' & SOAR & 147& 5 &  - &  -   	&  7202.74810 & 	21.6	& 142  &  Amor  &7.82 & - \\
2015~RF$_{2}$	& 09/28/15   & 129  & 1.032  &  0.046 &  48.3-48.9 & r' & SOAR &101 &9& - &  -  	& 7293.55333  & 	24.1 	&  44 &	  Apollo  & 6.59& - \\
2016~AG$_{166}$	& 01/19/16 & 110  &  1.060 & 0.088  & 29.0-28.9  & VR & DCT &40&5-10&	- & -	& 7406.96436  & 	 	24.0 & 47  &	  Apollo  &7.50 & - \\
2016~AU$_{65}$	&  01/19/16  & 117  &  1.146-1.145 & 0.164  &  8.0-8.2 & VR & DCT &90 &10& - & -    	& 7406.70597  & 	22.9	& 78  &	  Aten   &  11.57& -\\
2016~BE 	&  01/19/16  & 30  & 1.031  &  0.072 & 47.3  & VR & DCT &22&25-40& -  &   -  	& 7406.60448  & 	 23.7 	& 54  &	   Apollo &  6.19 & -\\
2016~BJ$_{15}$	& 02/08/16   & 81  &  1.071 &  0.091-0.092 & 20.4-20.3  & wh  & KP4 &156&20-30&  - &    -	& 7426.88096  & 	 23.3  	& 65  &	 Apollo   & 6.15& - \\
2016~CF$_{29}$	& 02/08/16  & 92  & 1.032-1.031  & 0.051  &  26.5-27.0 & wh  & KP4 &122&8-15&  - &   - 	& 7426.79262  & 	  24.9 	&  31 &	  Apollo  & 7.22& - \\
2016~CL$_{29}$	&  02/08/16  & 82  &  1.046 &  0.065 & 24.2-24.3  & wh  & KP4 &80&15 & - &  -  	& 7426.63062  & 	  24.6 	& 35  &	  Apollo  & 7.79 & -\\
2016~EQ$_{1}$	&  03/16/16  & 50  &  1.028   &  0.033  &  5.7  & VR  & DCT &15&1-4&  - &    -	&   7463.93605 & 	   26.3	&  16 &	 Apollo   & 4.73 &  9.915  \\
2016~FX$_{2}$	&  03/22/16  &  71 &  1.033  & 0.058   &  49.2-49.1  & VR  & DCT &41&20&  - &    -	&  7469.64692 & 	 23.7  	& 54  &	  Apollo  & 4.52 & 8.014  \\
2016~GB$_{222}$	&  04/19/16  & 145  &  1.017 &  0.016 &  37.4-38.1 & VR  & DCT &68&5&  - &    -	&  7497.81978 & 	 26.8  	& 12  &	 Apollo   & 5.84& - \\
2016~GV$_{221}$	&  04/19/16  & 201  & 1.013  & 0.026  & 69.6-70.3  & VR  & DCT &105&2-10&  - &    -	& 7497.63660  & 	 24.9  	& 31  &	 Apollo   & 5.49 & 10.590  \\
\hline\hline
\textit{Tumblers} &		& & & & & & &	&			&	 	&&	&&&	\\
2013~YG 	& 01/03/14 & 90  & 1.032  & 0.053  & 23.3-23.4  & wh & KP4 &97&15&	0.2921 &   -	& 6660.82729  & 	 25.4 	& 24   &	 Aten  &  5.29& 6.637 \\
2014~DJ$_{80}$	& 03/26/14 &  120 &  1.043 &  0.059 &  38.3-38.6 & r' &SOAR &166&45&	-&   -	& 6742.50997   & 	 26.3 	& 16   &	 Aten  & 4.41 & 5.527 \\
2015~CG	        & 02/10/15   & 398  &  1.002 &  0.020 &  38.5-40.1 & VR & DCT &134&3& 0.0353 &  -   	&  7063.91241 & 	25.6 	&22   &	 Apollo    & 6.58& - \\
2015~HB$_{177}$	   &  05/12/15   &  114  &  1.023 &  0.035 & 67.3-67.7  &    VR & DCT &46&3& -     &	-&  7154.93442 & 	24.6 	& 35   &	 Apollo    & 6.61 & 11.795 \\
2015~LJ 	& 08/18/15 & 136  & 1.066  & 0.061  & 26.8  & VR & DCT&99&15-25& 0.1875	 &  -	& 7252.72284  & 	24.7 	&  34 &	  Amor  &  4.36 & 8.404\\
2016~FA 	&  03/16/16  & 199  & 1.016  & 0.030  &  44.8-44.7 & VR & DCT &57&3-5&  -  &    -  	&  7463.97424 & 	25.2 	&  27 &	 Aten   &  5.96 & 6.918\\
2016~RD$_{34}$ & 09/15/16 & 190 & 1.008& 0.0087 &72.8-72.7  & VR &DCT &79&2-5&0.0230		&	-		& 7646.84794	&   27.6 &  8 &	Amor  & 3.83& 4.233 \\
2017~EE$_{3}$ & 03/09/17 & 87 & 1.012& 0.024& 38.0-38.1& VR & DCT&30&5&	0.1050	&		-	&	7821.73596 &   26.0 &  18 &	Apollo  & 5.64& - \\
2017~HU$_{49}$ & 05/14/17& 274 & 1.022 & 0.014 &  36.5-37.0 & VR & DCT&135&3-2&	0.62	&	-		& 7887.65057	&   26.5  &  14 &	Aten  & 4.49& 4.473 \\
2017~QW$_{1}$   &08/21/17 &121 &1.047 & 0.036 & 15.4-15.5 & VR& SOAR& 68 &15&		0.0987&		- 	& 7986.74562  &   26.2 & 17  &	Aten  &4.82 & 5.171 \\ 
\enddata
\tablenotetext{a}{UT-date, distance sun-object (r$_{h}$), and distance earth-object ($\Delta$) and phase angle ($\alpha$) are summarized. The filter (details in \citet{Thirouin2016}), telescope (Tel) and number of images ($Nb_{im}$), rotational period (Rot. P. in hour), relative amplitude ($\Delta$m), and the Julian Date ($\varphi_{0}$) for the zero phase are presented. No light-time correction applied. Absolute magnitude (H), and an estimate of the NEO diameter (D) with 20$\%$ as albedo are also reported. Exposure time (t$_{exp}$), and duration of the observing block ($\Delta T$) for each NEO is indicated.\\
$^{b}$: Two other lightcurves have been published for this object by \citet{Warner2015SM143} suggesting a rotational period of 31~h and by \citet{Kikwaya2018} with a periodicity of 1~h. For the purpose of our work, we use the MANOS result reported here. \\
$^{c}$: Two lightcurves are reported for this object (Section~\ref{sec:flat}).}

\end{deluxetable}
\end{longrotatetable}

\clearpage

\begin{deluxetable*}{lccccccc}
\tablecaption{\label{Tab:candidates} Most suitable targets for a robotic/human mission$^{a}$.}
\tablecolumns{8}
\tablenum{2}
\tablewidth{0pt}
\tablehead{
\colhead{NEO} &
\colhead{H} &
\colhead{Diameter [m]} &
\colhead{Rot. period [h]} & \colhead{Vis. Spec.} & \colhead{$\Delta$$_{v}$$^{SH}$} &
\colhead{$\Delta$$_{v}$$^{NHATS}$}&
\colhead{Start next optical Window}}
\startdata
2016~DK	& 22.4	 & 98 &	1.30 &  no	&	11.34	&	-	 & - \\ 
2017~QX$_{1}$	& 24.8	 & 32 &	 1.34	&  yes	&	5.53	&	-	 & - \\ 
2016~NK$_{39}$	& 23.9	 & 49 &	 1.46	&  no	&	5.77	&	11.025	 & 2023/05 \\ 
2016~PP$_{27}$	& 23.6	 & 56 &	 1.55	&  yes	&	7.12	&	-	 & - \\ 
2014~HS$_{184}$	&23.3 &65&2.02 &  yes	&5.72 &	- 	&  -\\ 
2010~MR	& 18.6	 & 566 &	 2.42	&  no	&	6.80	&	-	 & - \\ 
2016~NG$_{38}$	& 25.1	 & 28 &	 2.47	&  no	&	5.98	&	-	 & - \\ 
2015~AA$_{44}$	& 23.9	 &  49  &	 $>$1	&  no	&	5.68	&	-	 & -  \\ 
2015~QA	& 22.9 &  78 &	 $>$1	&  no	&	6.56 &	-	 & -  \\ 
2015~PT$_{227}$	& 23.9 &  49 &	 $>$1	&  yes	&	6.29 &	-	 & -  \\ 
2016~HN$_{2}$	& 23.5	 & 59  &	 $>$1	&  yes	&	6.08	&	-	 & -  \\ 
2016~HP$_{3}$	& 23.7	 &  54  &	 $>$1	&  no	&	6.46	&	-	 & -  \\ 
2016~JE$_{18}$	& 26.3	 &  16  &	 $>$1	&  no	&	5.88	&	-	 & -  \\ 
2017~EK$_{3}$	& 26.3	 & 16 &	 $>$1.5	&  no	&	5.87	&	8.840	 & none \\ 
2015~ET	& 26.7	 &  13  &	 $>$2	& no	&	6.48	&	-	 & -  \\ 
2013~VY$_{13}$	& 21.2	 &  171  &	 $>$2	&  yes	&	6.80	&	-	 & -  \\ 
\textit{\textbf{2013~YS$_{2}$}}	& \textit{\textbf{23.3}} & \textit{\textbf{65}}  & 	\textit{\textbf{$>$2}}  &  \textit{\textbf{yes}}	& \textit{\textbf{4.77}} & \textit{\textbf{10.346}} & \textit{\textbf{ 2020/09}} \\
\textit{\textbf{2014~FA$_{7}$}}	& \textit{\textbf{26.7}}	 &  \textit{\textbf{13}}  &	 \textit{\textbf{$>$2}}	&  \textit{\textbf{yes}}	&	\textit{\textbf{5.17}}	&	\textit{\textbf{7.232}}	 & \textit{\textbf{2032/09}}  \\ 
2017~QU$_{17}$	& 26.1	 & 17 &	 $>$2	&  no	&	6.20	&	-	 & - \\ 
2013~XX$_{8}$	& 24.4	 &  39  &	 $>$2.5	&  no	&	4.57	&	10.364	 & 2019/04  \\ 
 2014~WO$_{69}$	& 23.6	 &  56  &	 $>$2.5	&  yes	&	6.19	&	-	 & -  \\ 
2014~HK$_{129}$	& 21.1	 &  179  &	 $>$3	&  yes	&	6.26	&	-	 & -  \\ 
\enddata
\tablenotetext{a}{MANOS obtained spectra and lightcurves for two good spacecraft targets (italic/bold), but we also summarize all NEOs with a rotational period longer than 1~h. For completeness purposes, $\Delta$$_v$$^{SH}$ and $\Delta$$_v$$^{NHATS}$ following the \citet{Shoemaker1978} protocol and the NHATS parameters are summarized. The start of the next opportunity to observe these objects according to NHATS is also shown (\url{https://cneos.jpl.nasa.gov/nhats/}).  
}
\end{deluxetable*}
 
\clearpage

\clearpage

 \clearpage

 \clearpage

\clearpage

\appendix
\section*{Appendix A} \setcounter{section}{1}
Example of Lomb periodograms for objects reported in this work. 

\clearpage
\begin{figure*}
\includegraphics[width=8cm, angle=0]{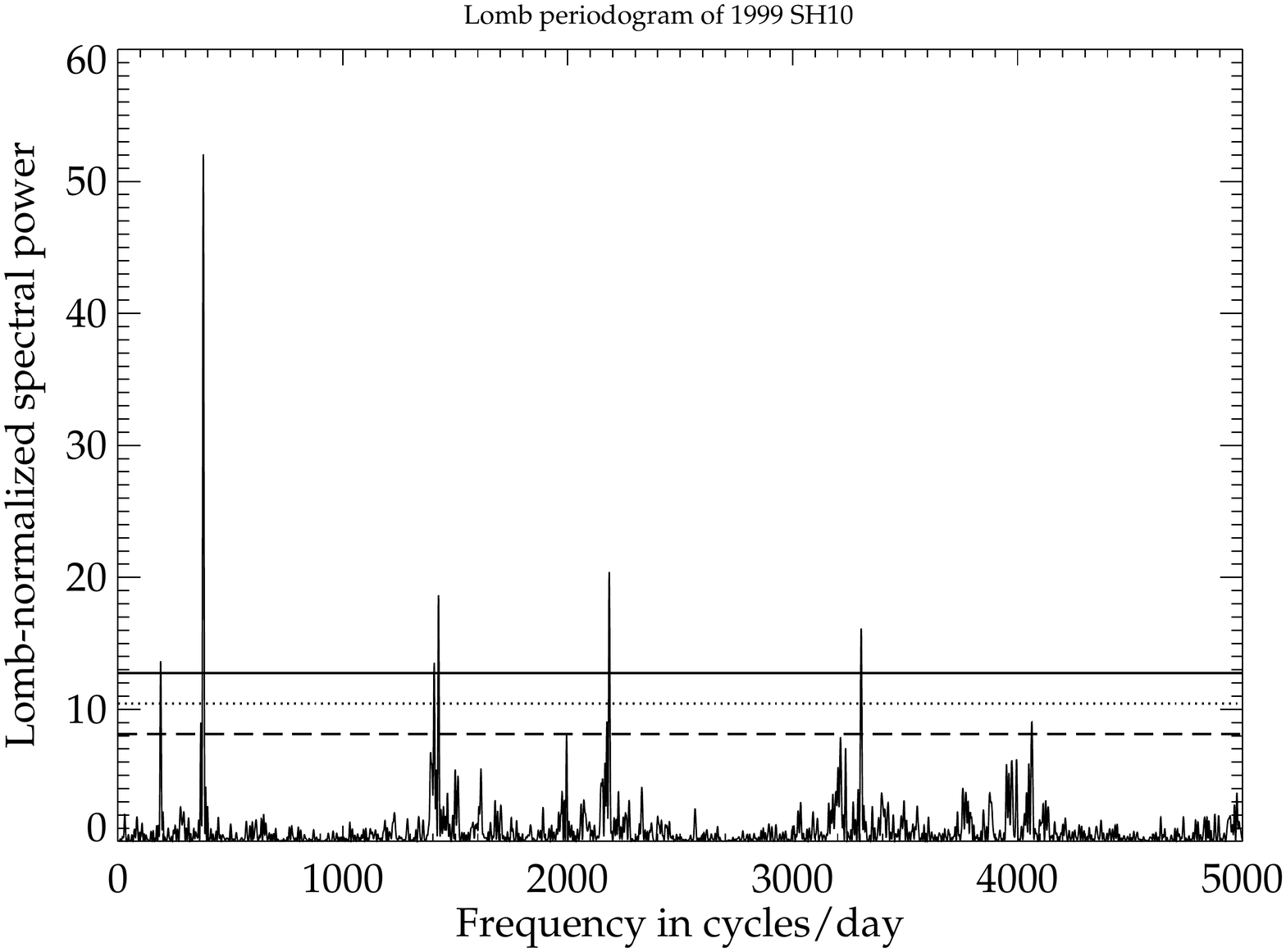}
\includegraphics[width=8cm, angle=0]{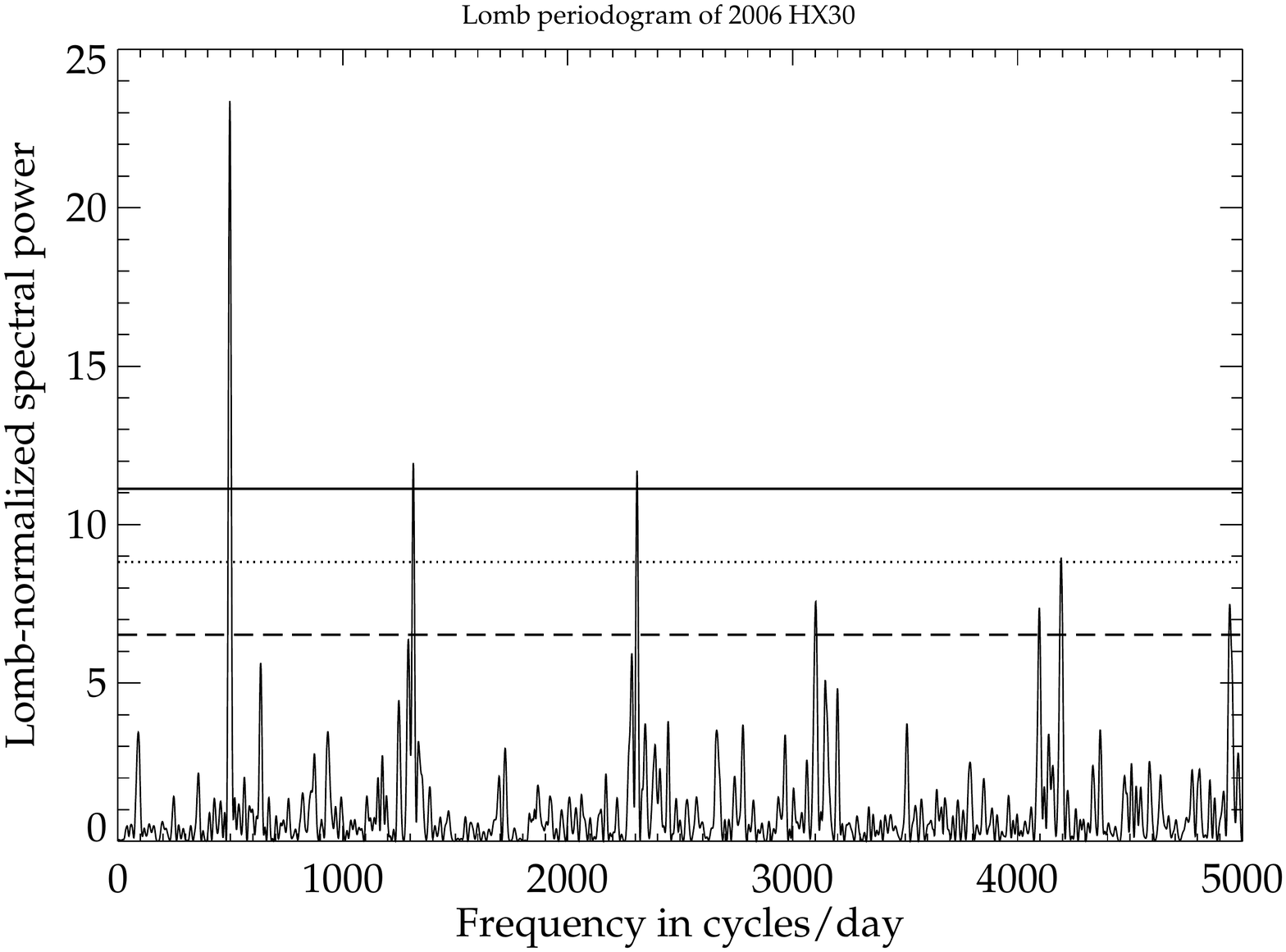}
\includegraphics[width=8cm, angle=0]{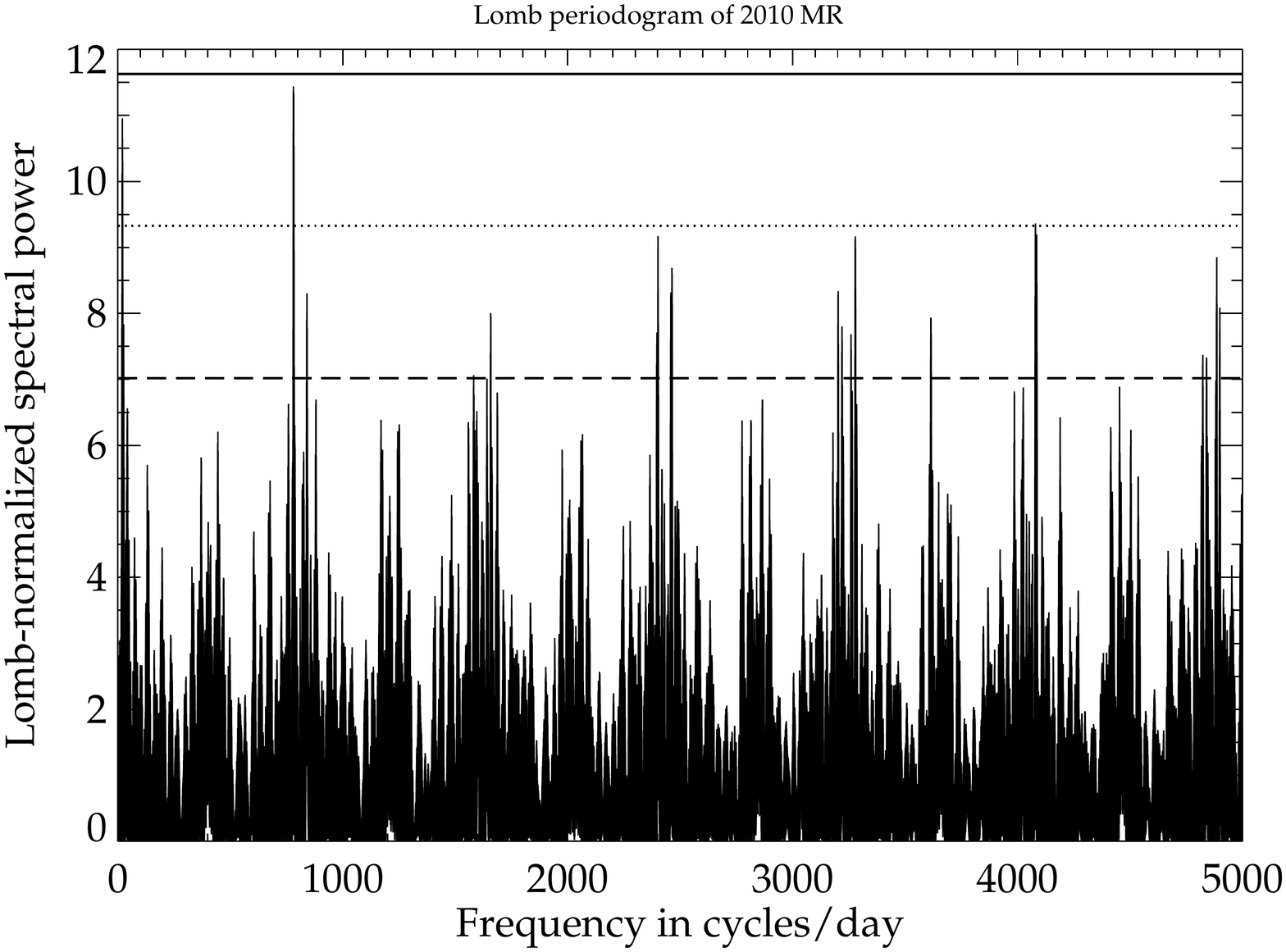}
\includegraphics[width=8cm, angle=0]{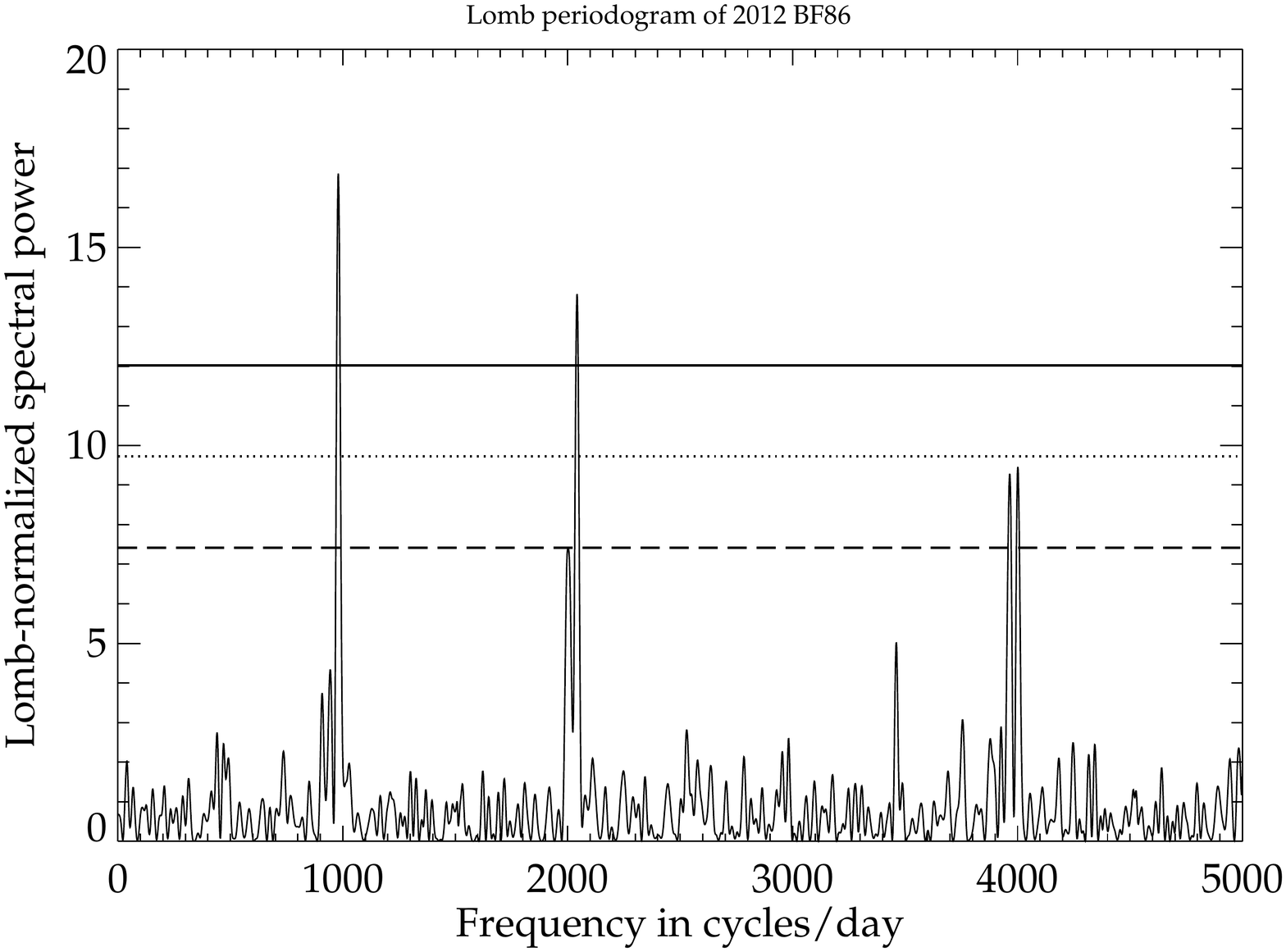}
\includegraphics[width=8cm, angle=0]{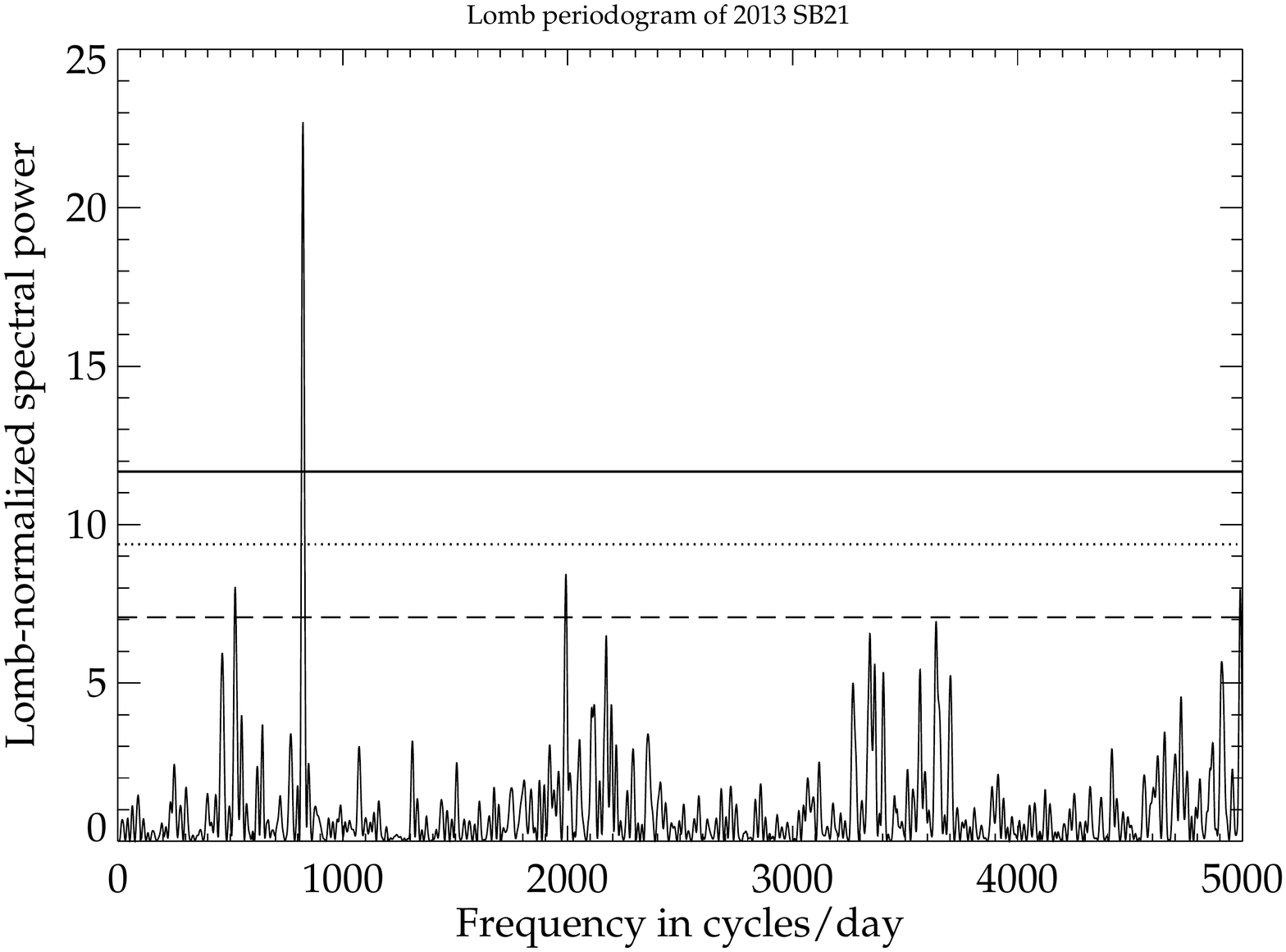}
\includegraphics[width=8cm, angle=0]{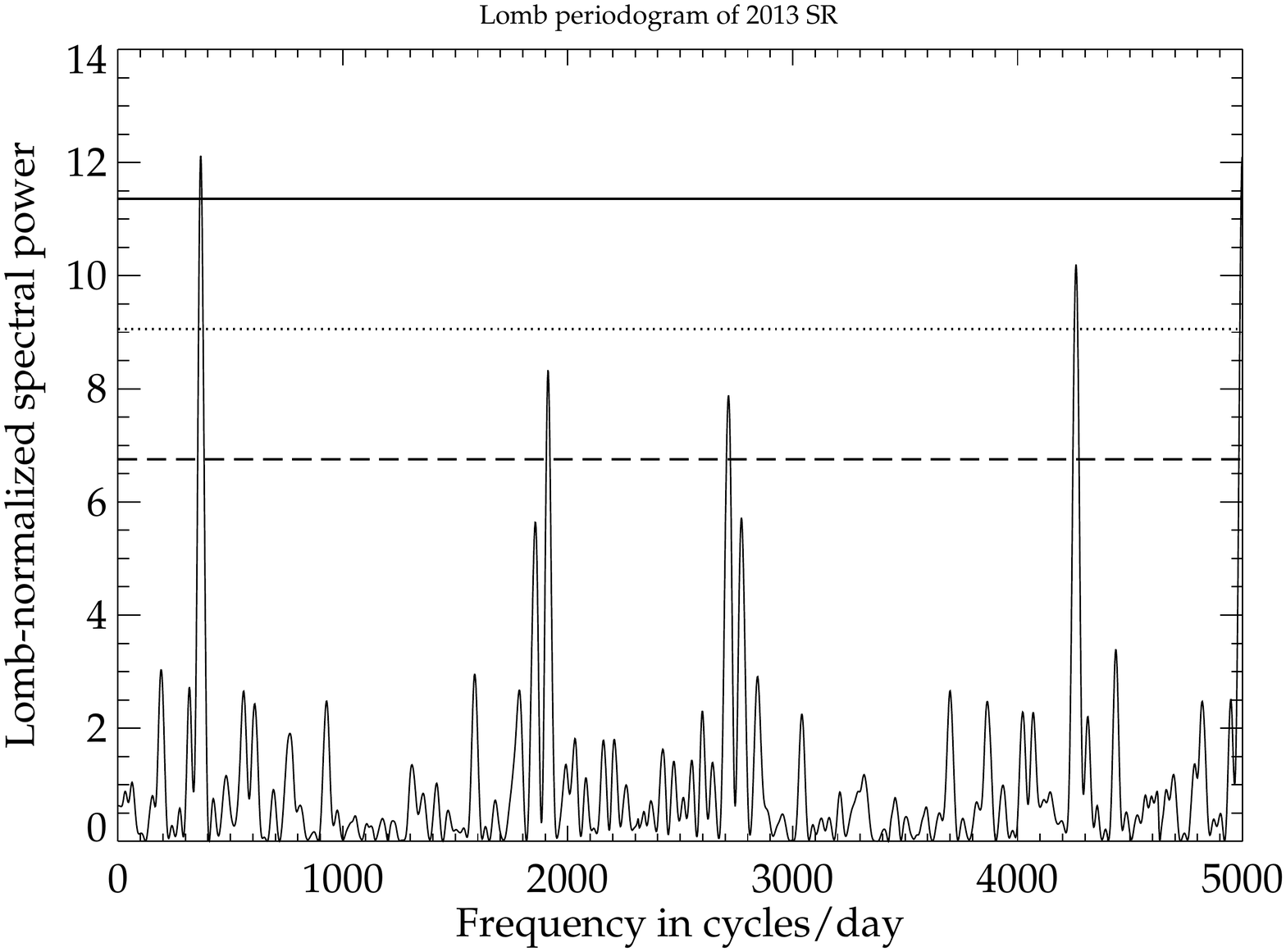}
\caption {Lomb periodograms are plotted with several confidence levels (continuous line: 99.9$\%$, dotted line: 99$\%$, and dashed line: 90$\%$).  }
\label{fig:Lomb1}
\end{figure*}   
\clearpage

\appendix
\section*{Appendix B}
\setcounter{section}{1}
Lightcurves of objects reported in this work. 

\begin{figure}
\includegraphics[width=24cm, angle=90]{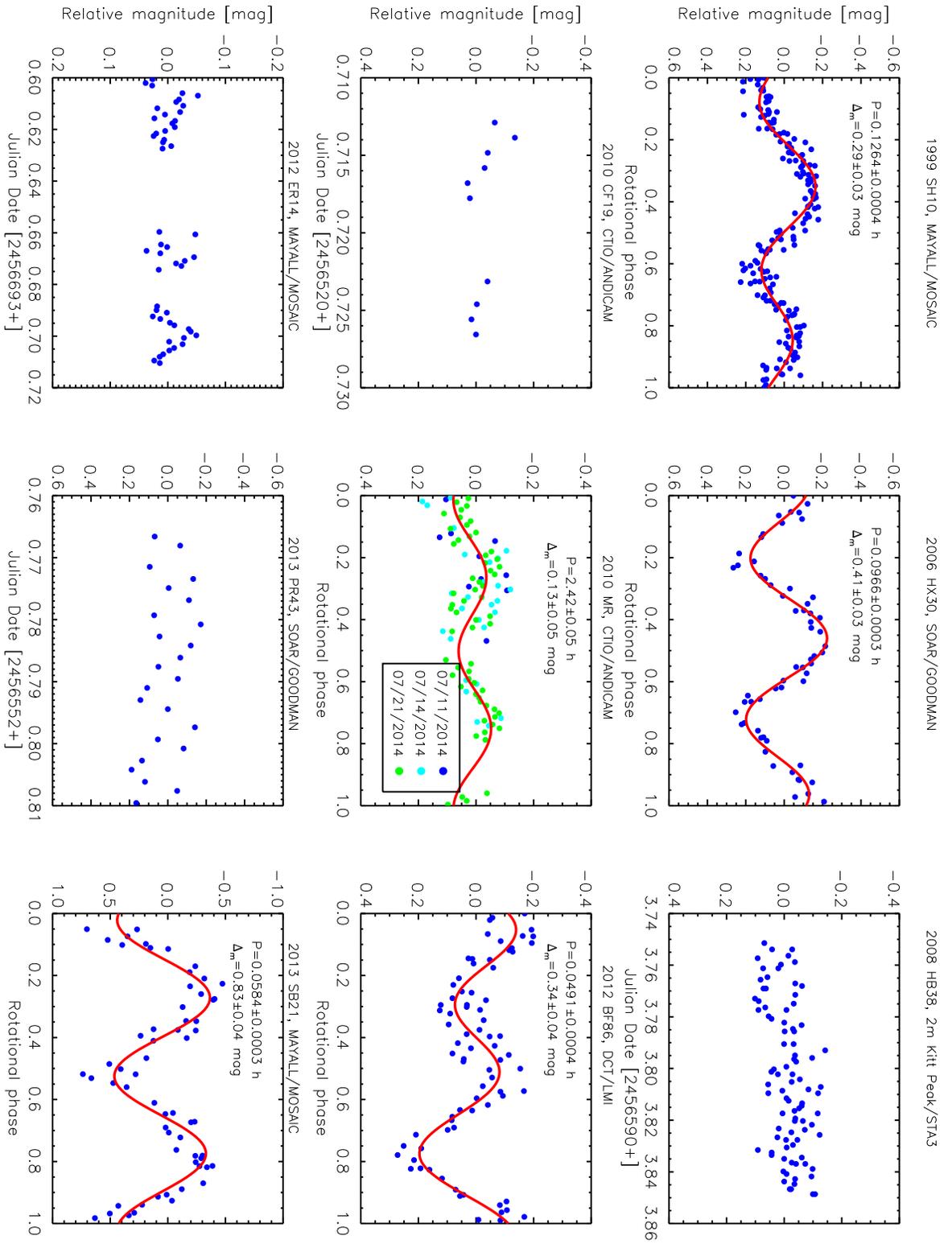} 
\caption {MANOS results are plotted.  }
\label{fig:LC1}
\end{figure}  
 
\clearpage

\begin{figure}
\includegraphics[width=24cm, angle=90]{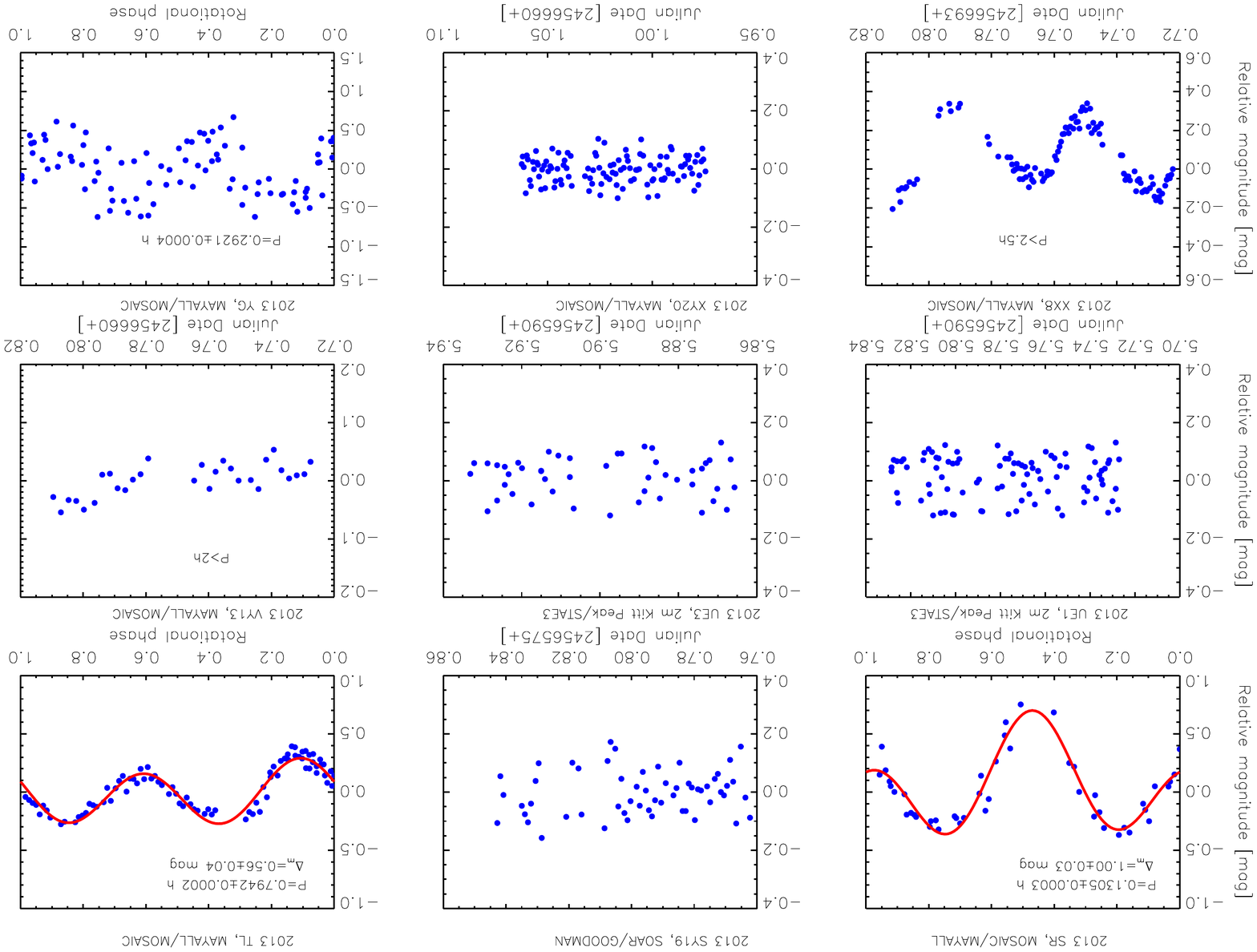} 
\caption {Continued}
\label{fig:LC2}
\end{figure}  
 
\clearpage
\begin{figure}
\includegraphics[width=24cm, angle=90]{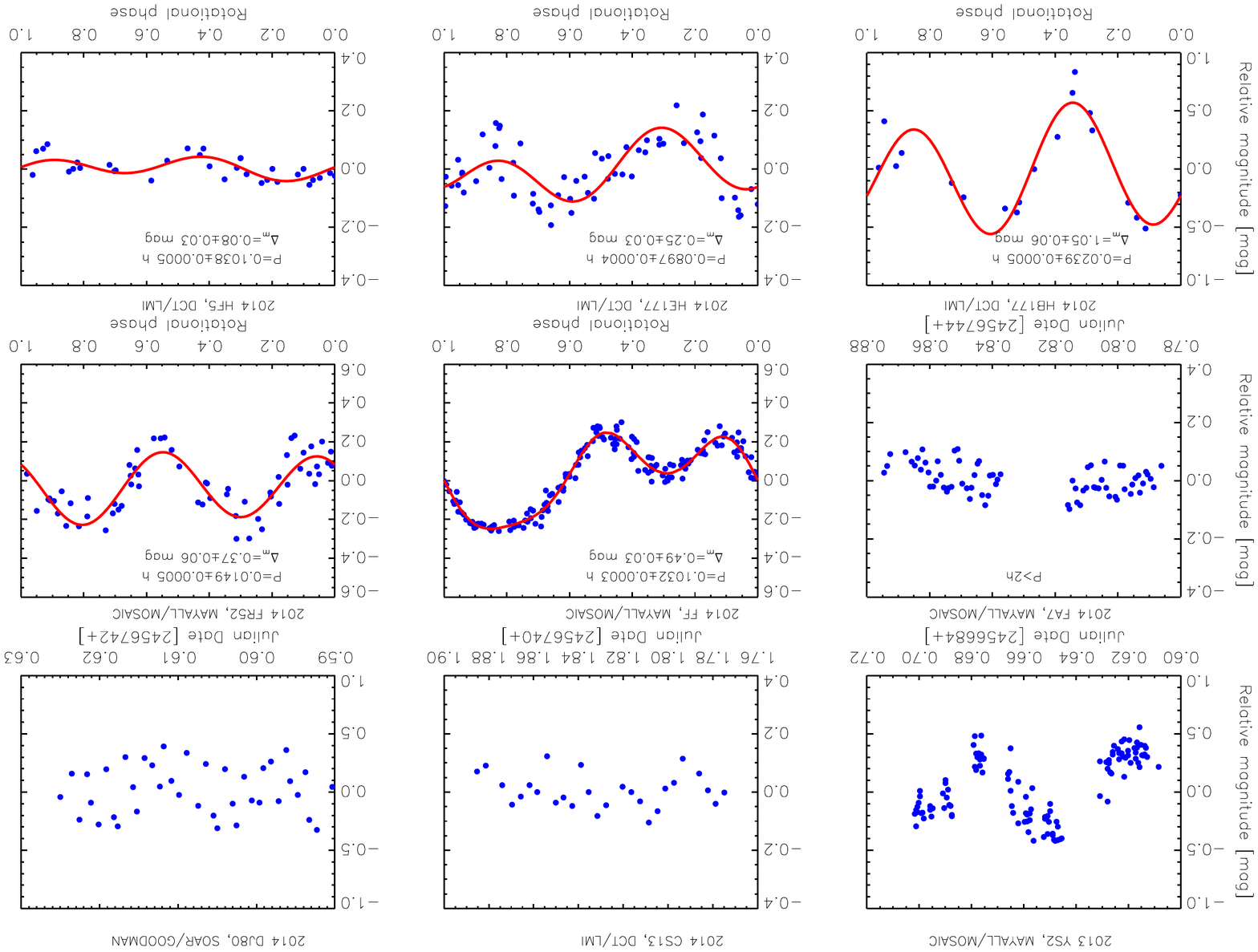} 
\caption {Continued }
\label{fig:LC3}
\end{figure}  
 
\clearpage
\begin{figure}
\includegraphics[width=24cm, angle=90]{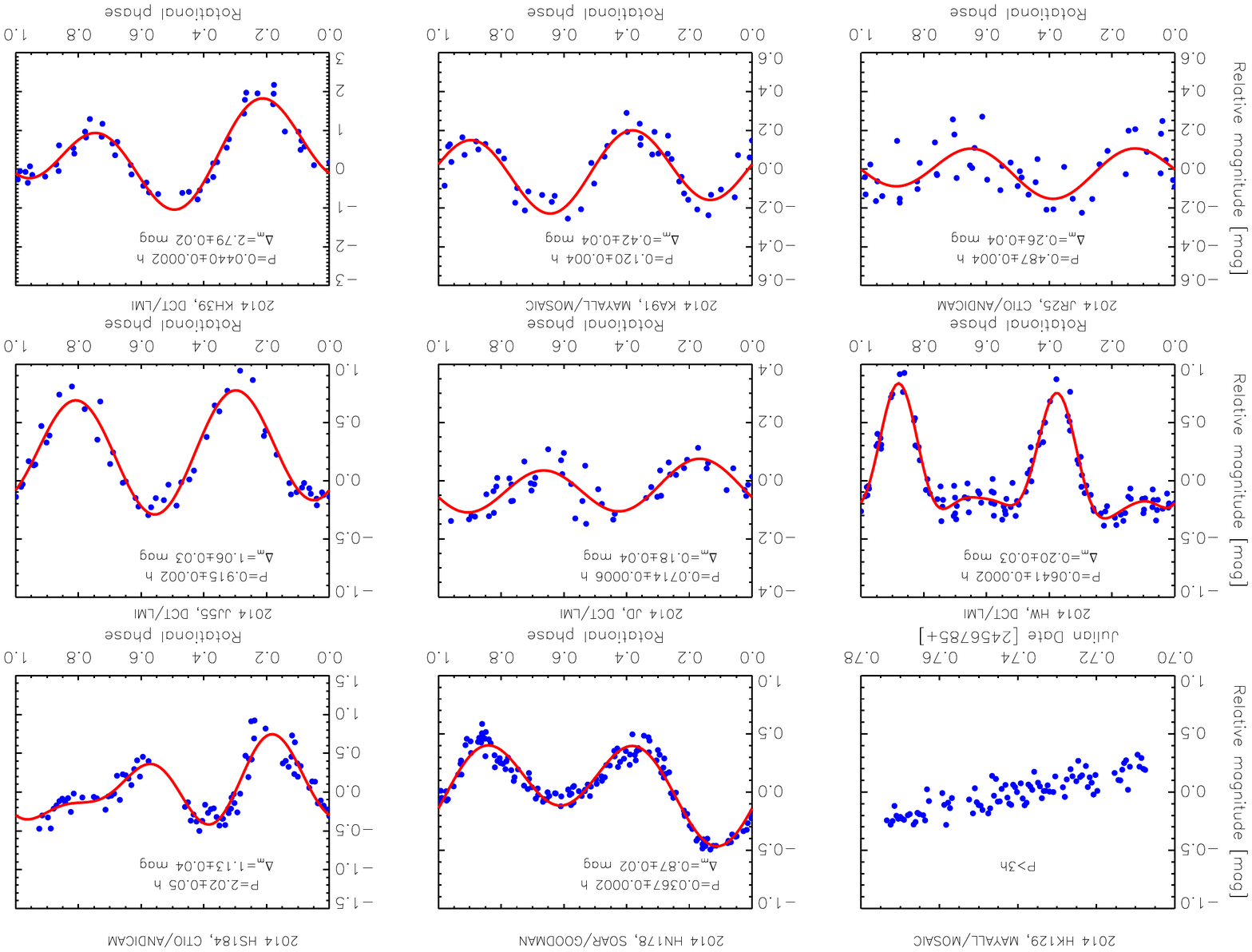} 
\caption {Continued }
\label{fig:LC4}
\end{figure}  
 
\clearpage
\begin{figure}
\includegraphics[width=24cm, angle=90]{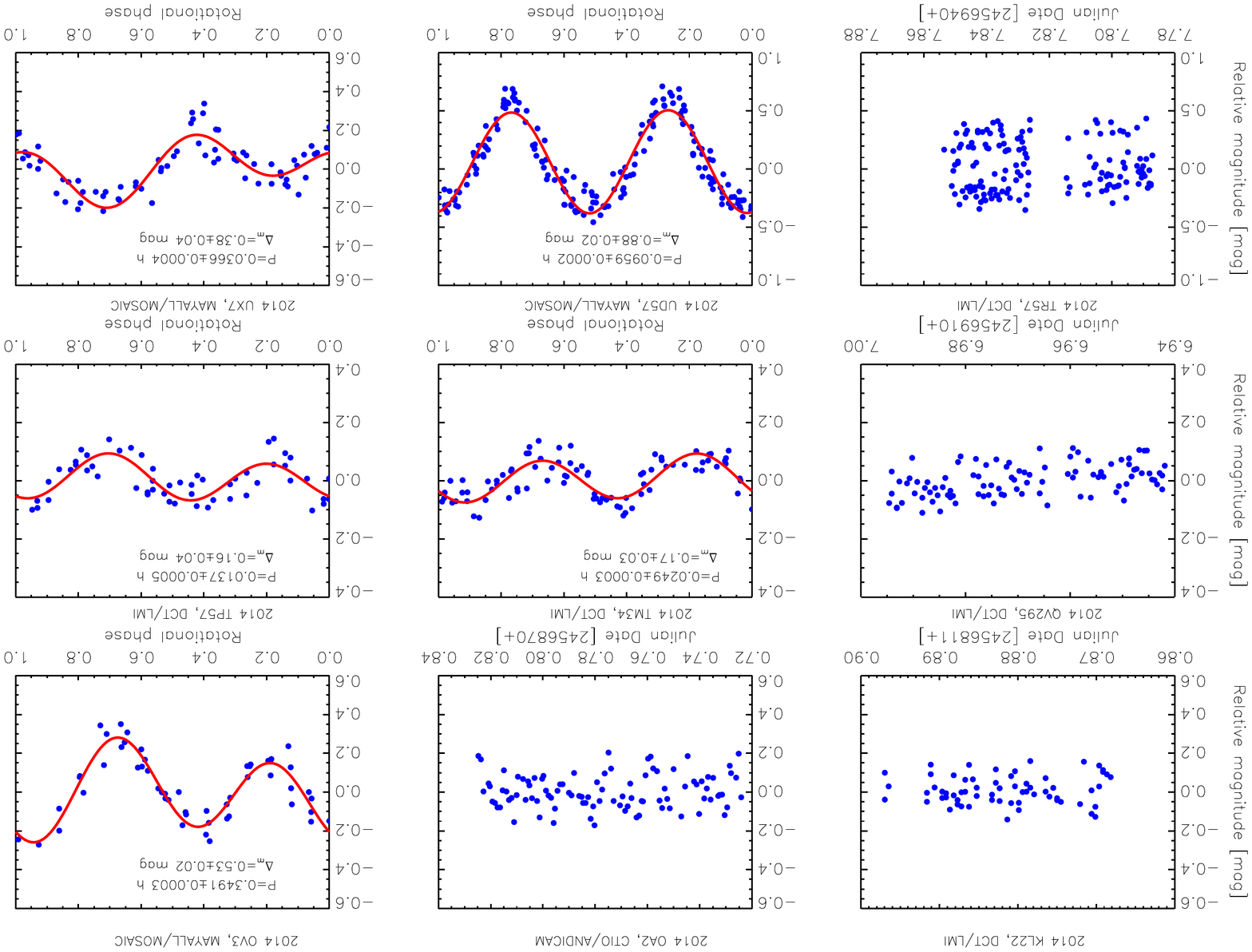} 
\caption {Continued }
\label{fig:LC5}
\end{figure}  
 
\clearpage
\begin{figure}
\includegraphics[width=24cm, angle=90]{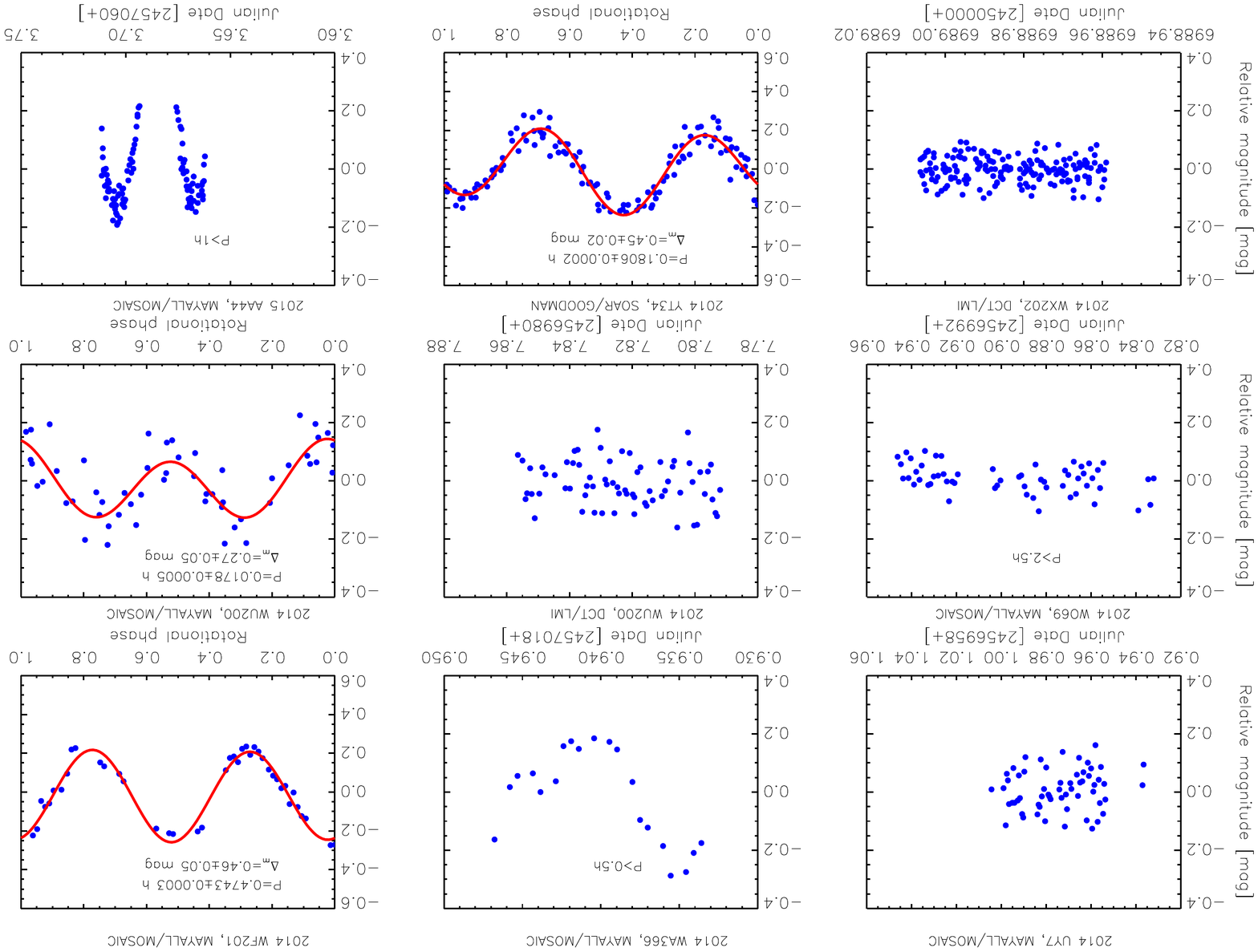} 
\caption {Continued }
\label{fig:LC6}
\end{figure}  
 
\clearpage

\begin{figure}
\includegraphics[width=24cm, angle=90]{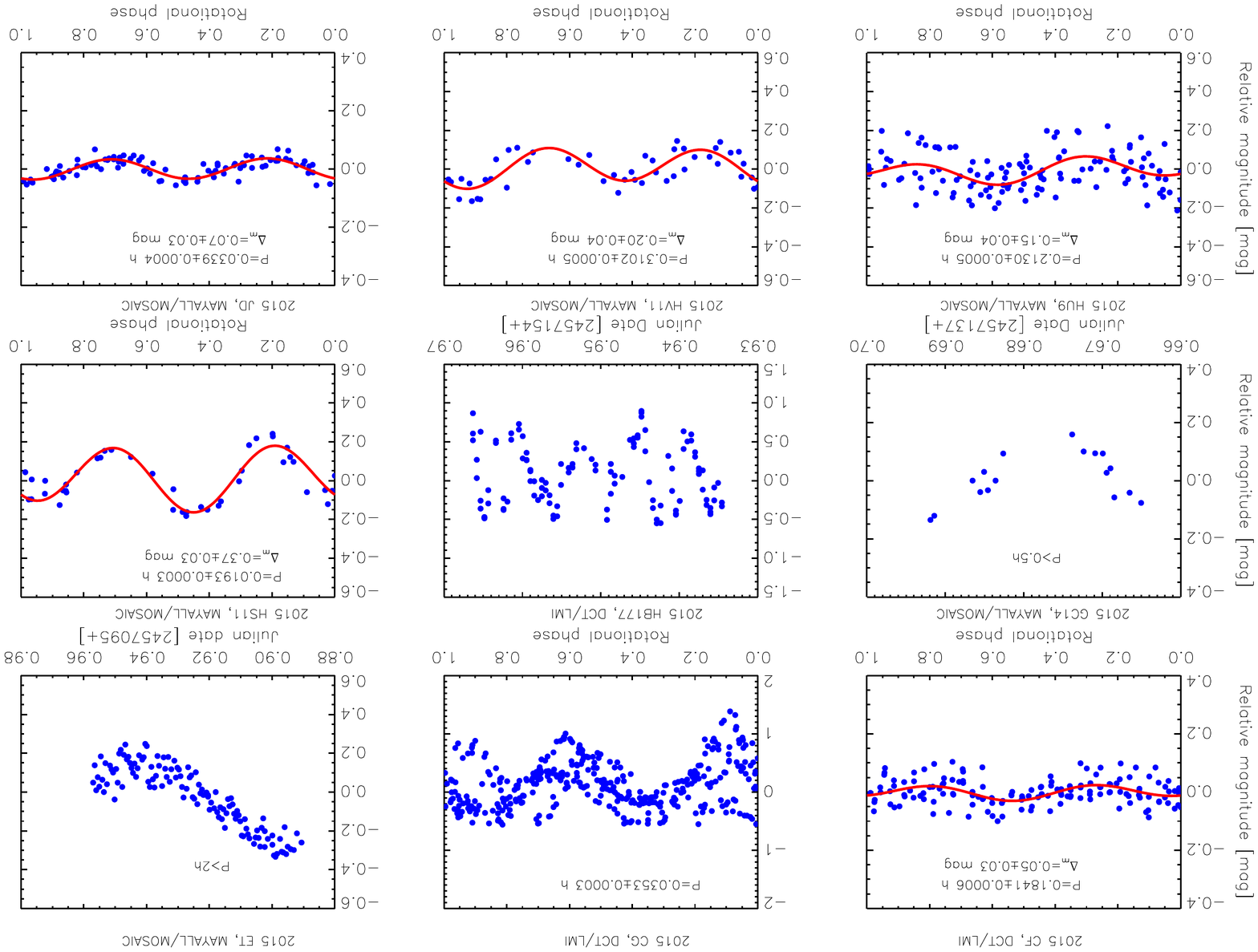} 
\caption {Continued }
\label{fig:LC7}
\end{figure}  

\clearpage

\begin{figure}
\includegraphics[width=24cm, angle=90]{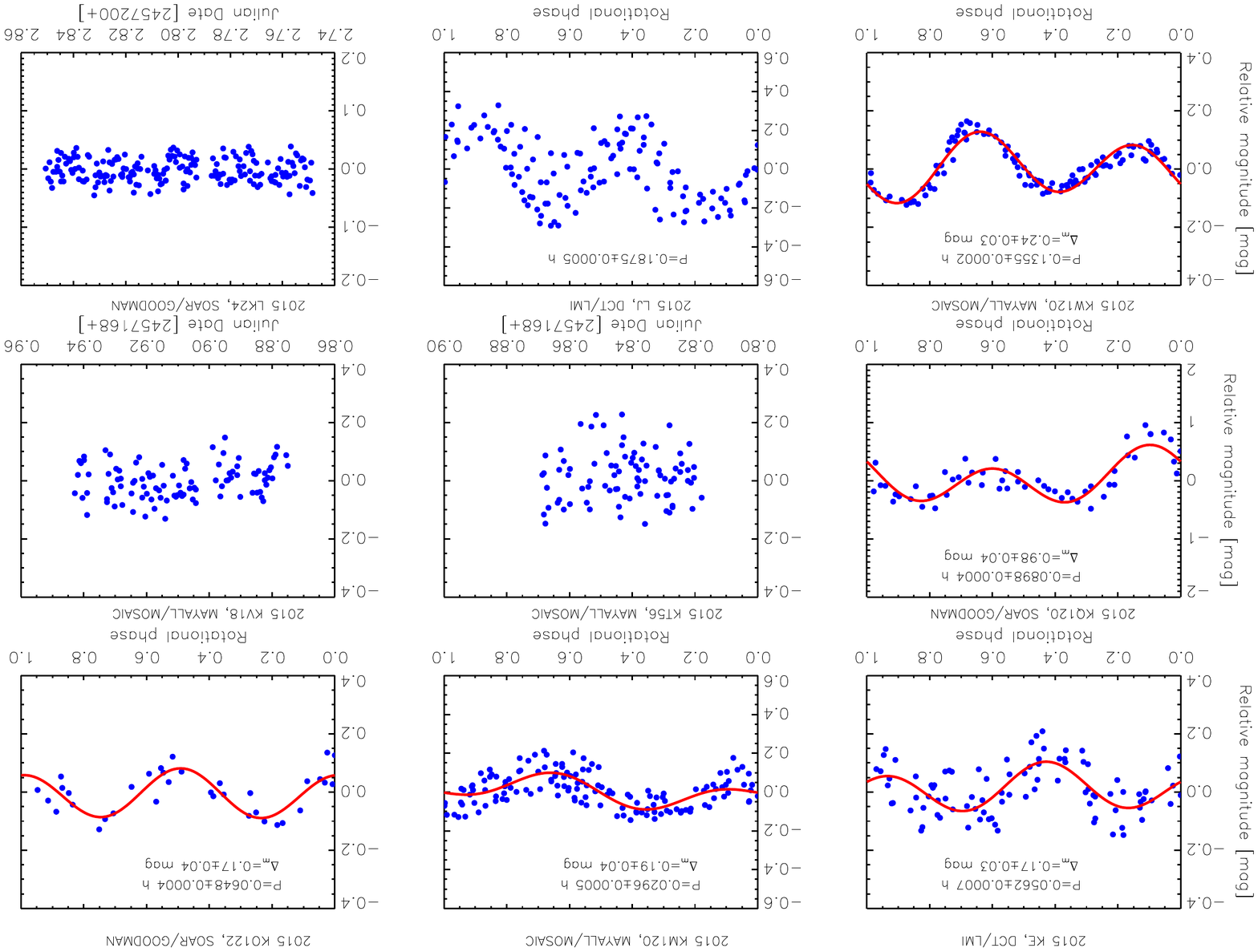} 
\caption {Continued }
\label{fig:LC8}
\end{figure}

\clearpage

\begin{figure}
\includegraphics[width=24cm, angle=90]{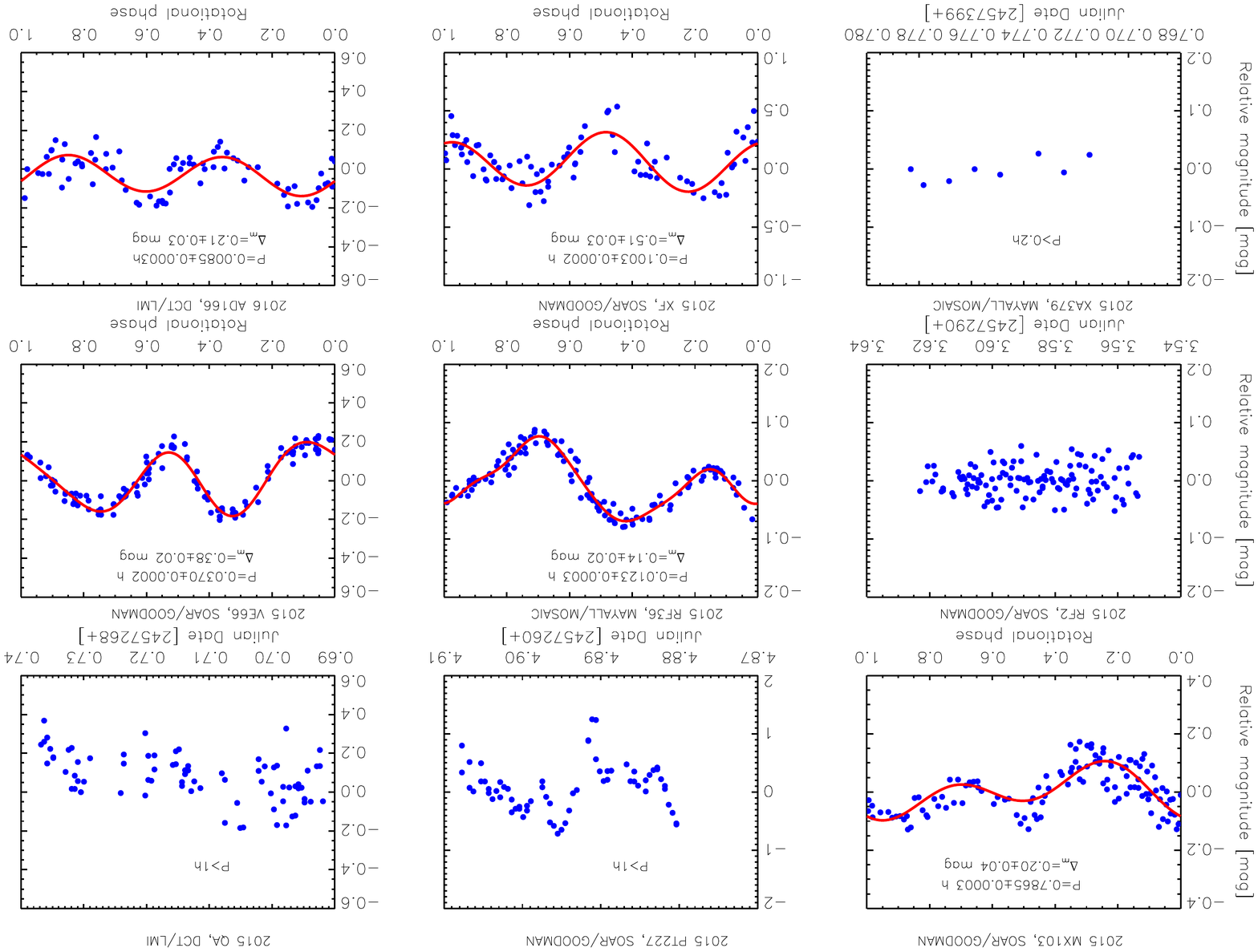} 
\caption {Continued }
\label{fig:LC9}
\end{figure}
 
\clearpage
\begin{figure}
\includegraphics[width=24cm, angle=90]{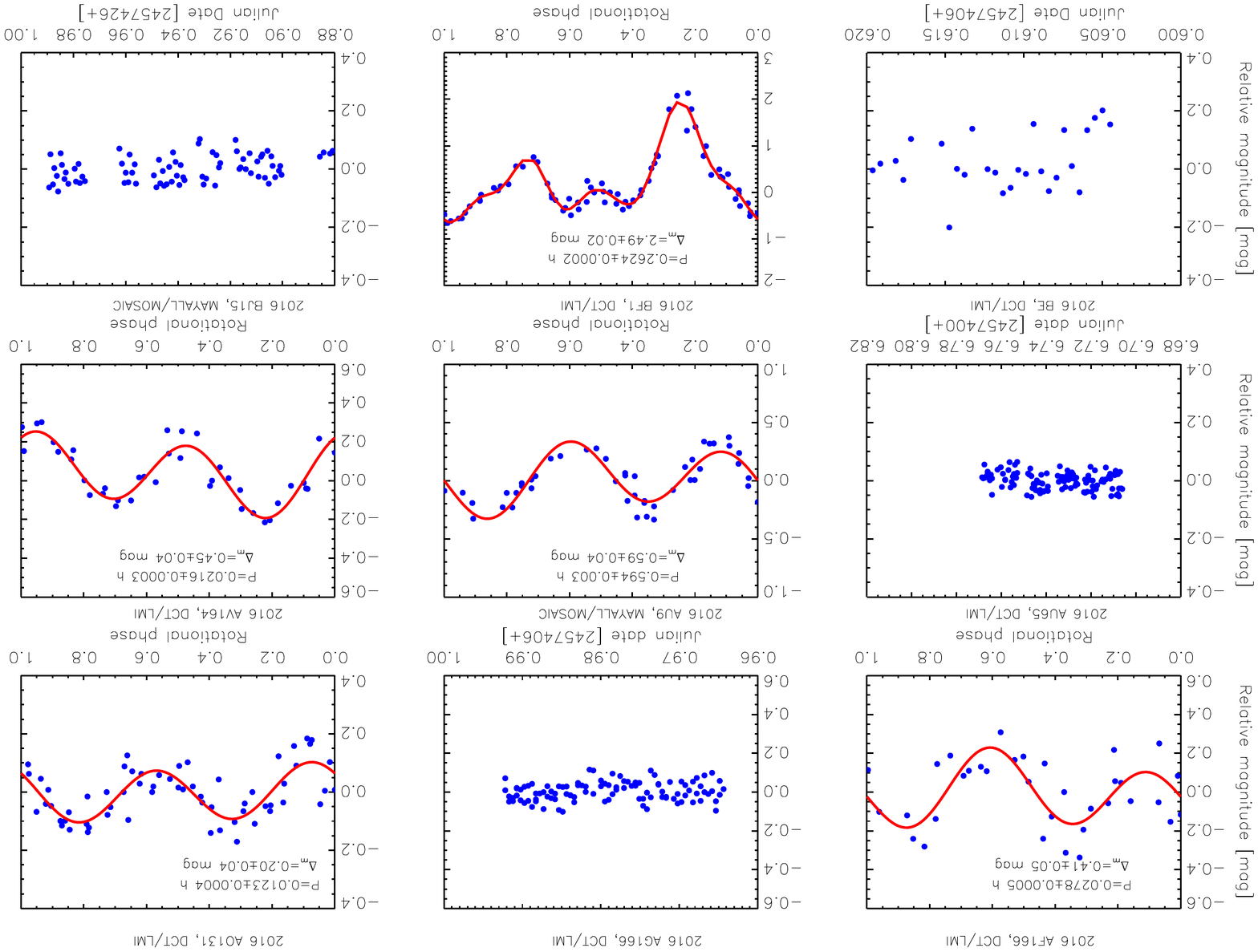} 
\caption {Continued }
\label{fig:LC10}
\end{figure}  
 
\clearpage
\begin{figure}
\includegraphics[width=24cm, angle=90]{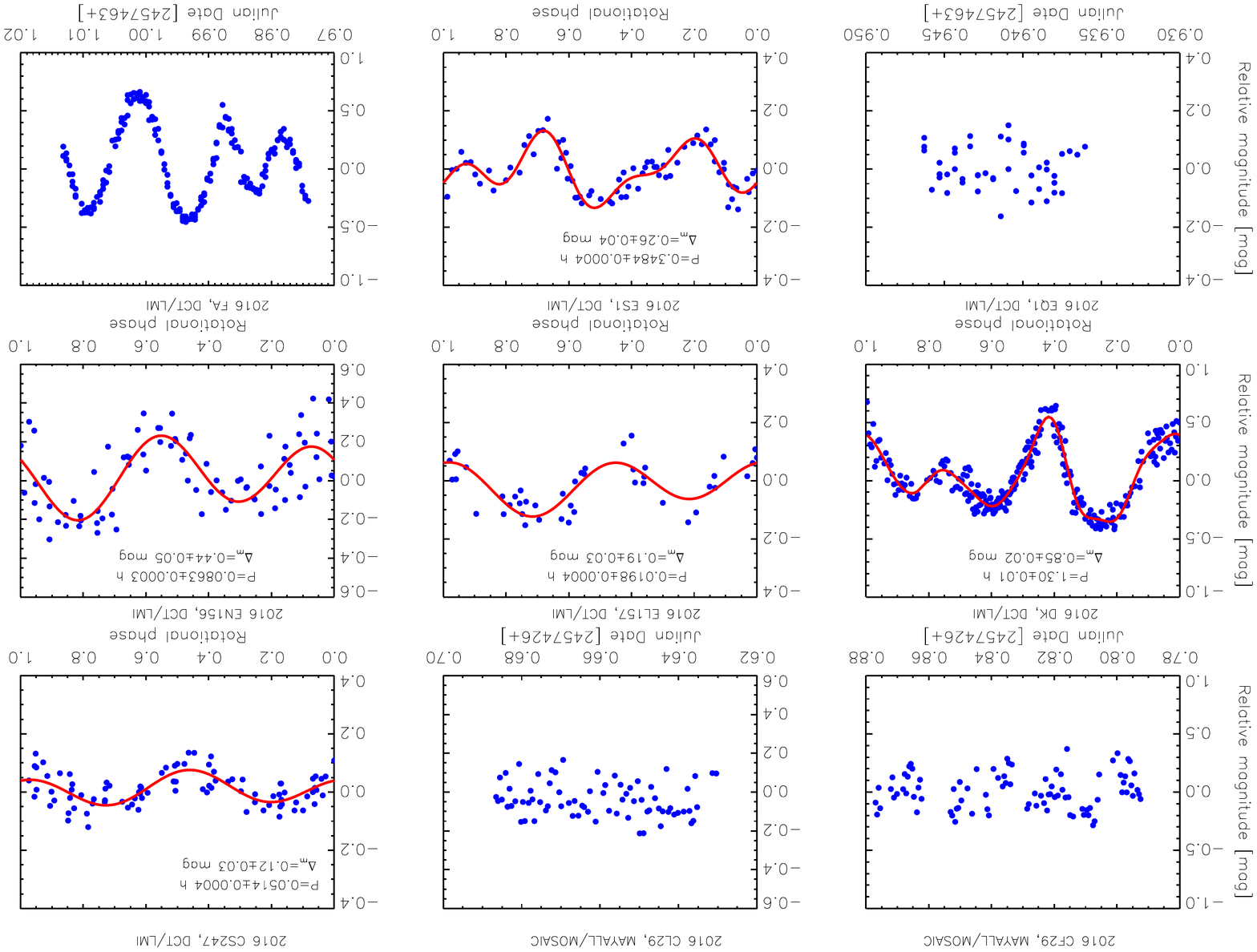} 
\caption {Continued }
\label{fig:LC11}
\end{figure}  
 
\clearpage

\begin{figure}
\includegraphics[width=24cm, angle=90]{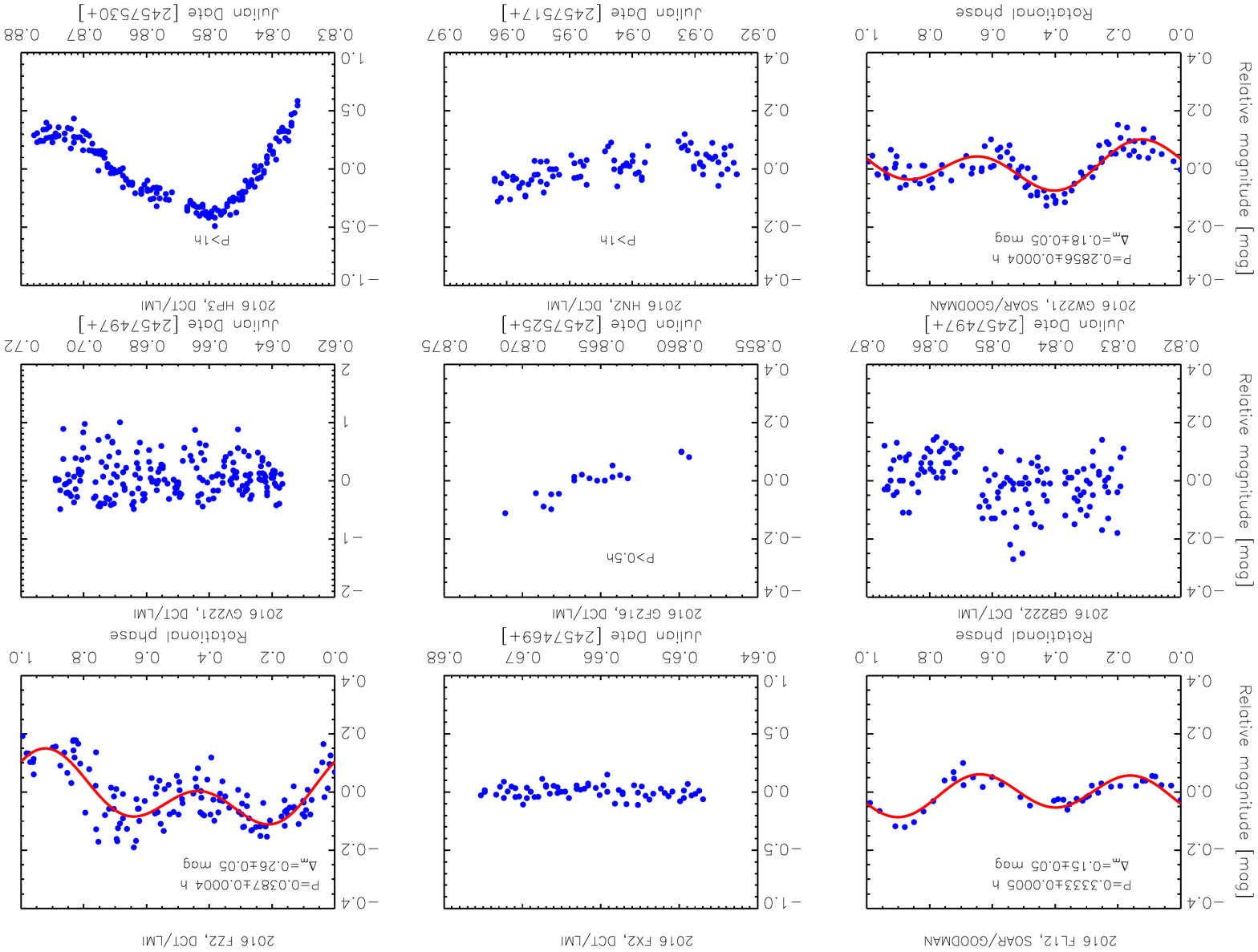} 
\caption {Continued } 
\label{fig:LC12}
\end{figure}  

\clearpage

\begin{figure}
\includegraphics[width=24cm, angle=90]{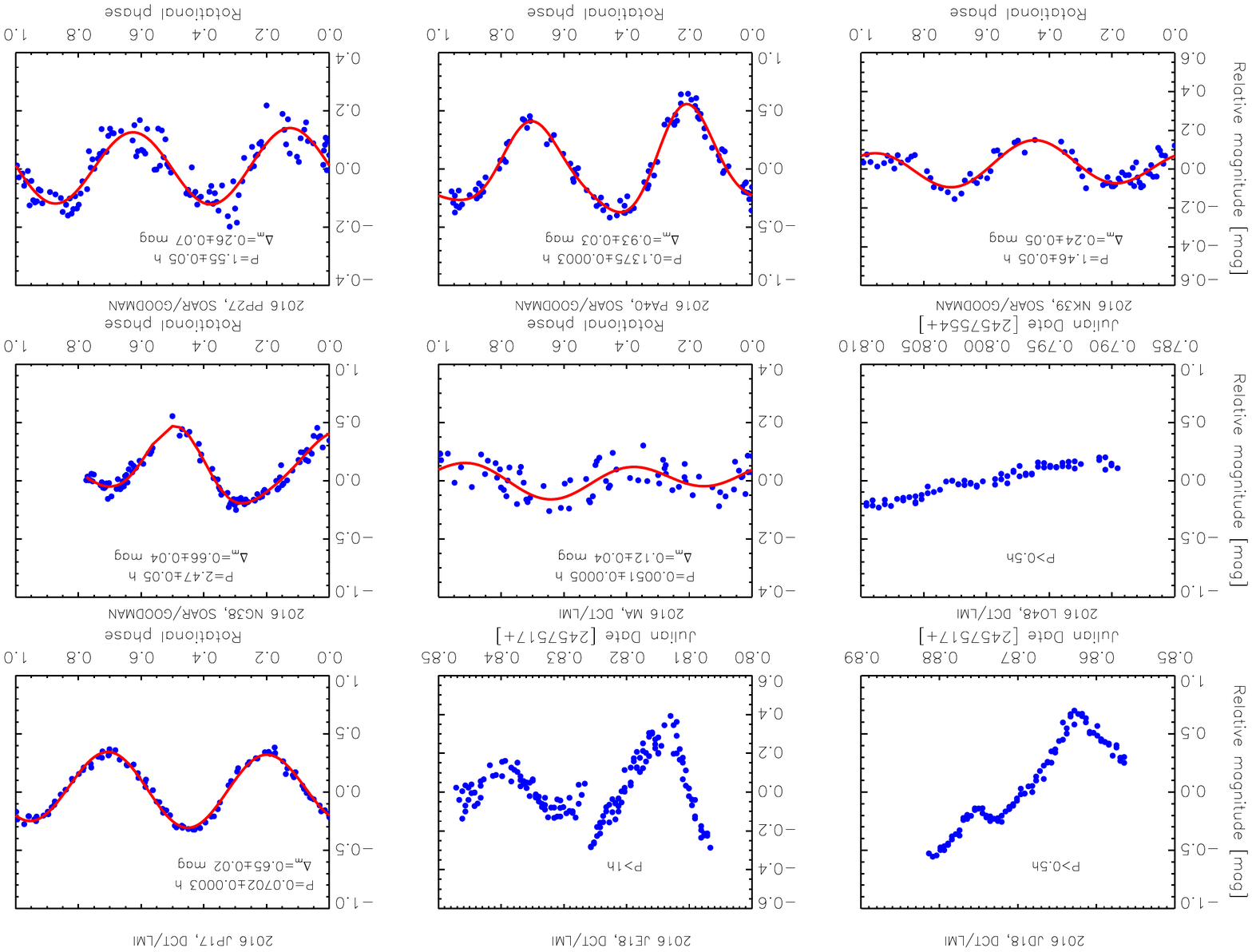} 
\caption {Continued }
\label{fig:LC13}
\end{figure}

\clearpage

\begin{figure}
\includegraphics[width=24cm, angle=90]{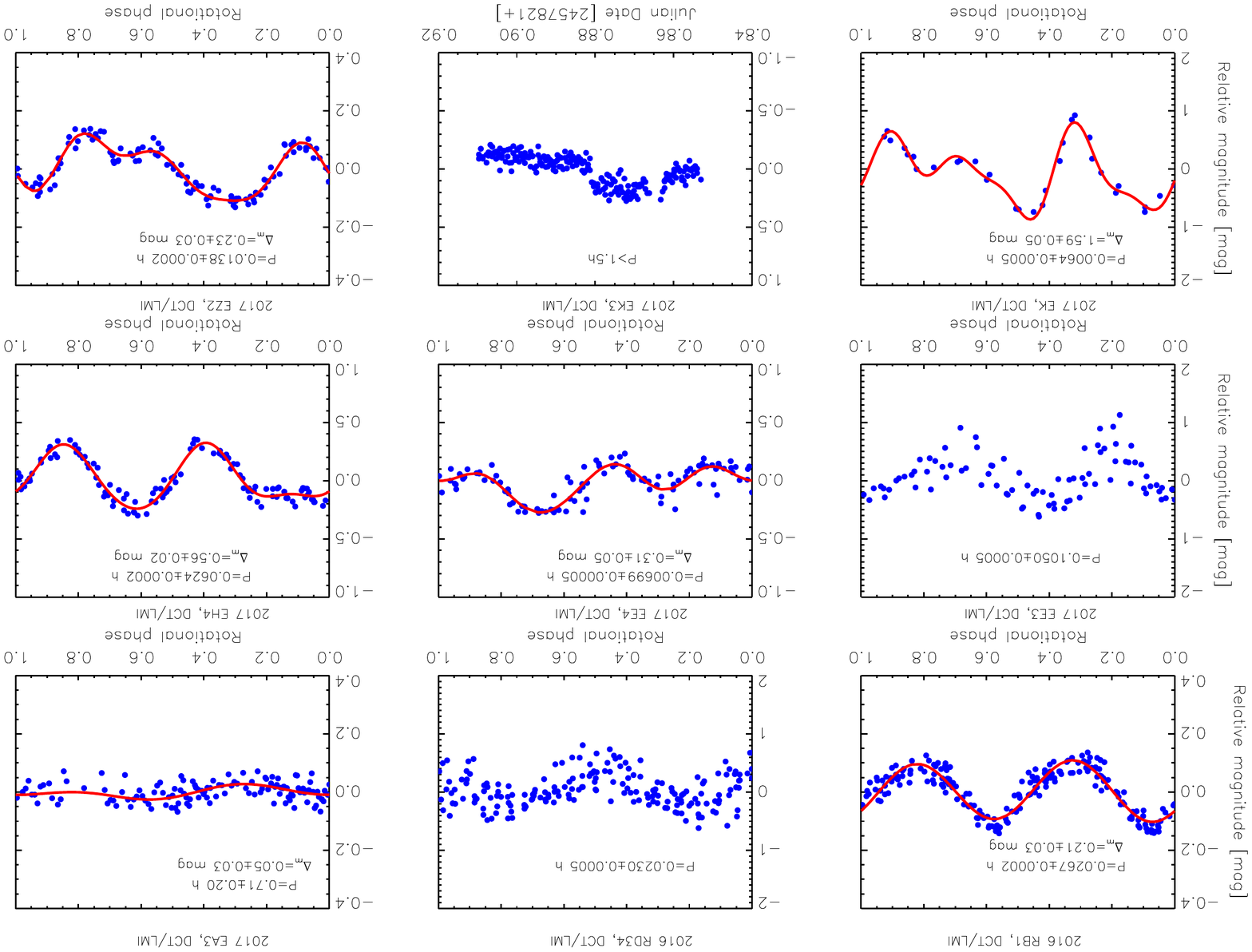} 
\caption {Continued }
\label{fig:LC13}
\end{figure}

\clearpage

\begin{figure}
\includegraphics[width=24cm, angle=90]{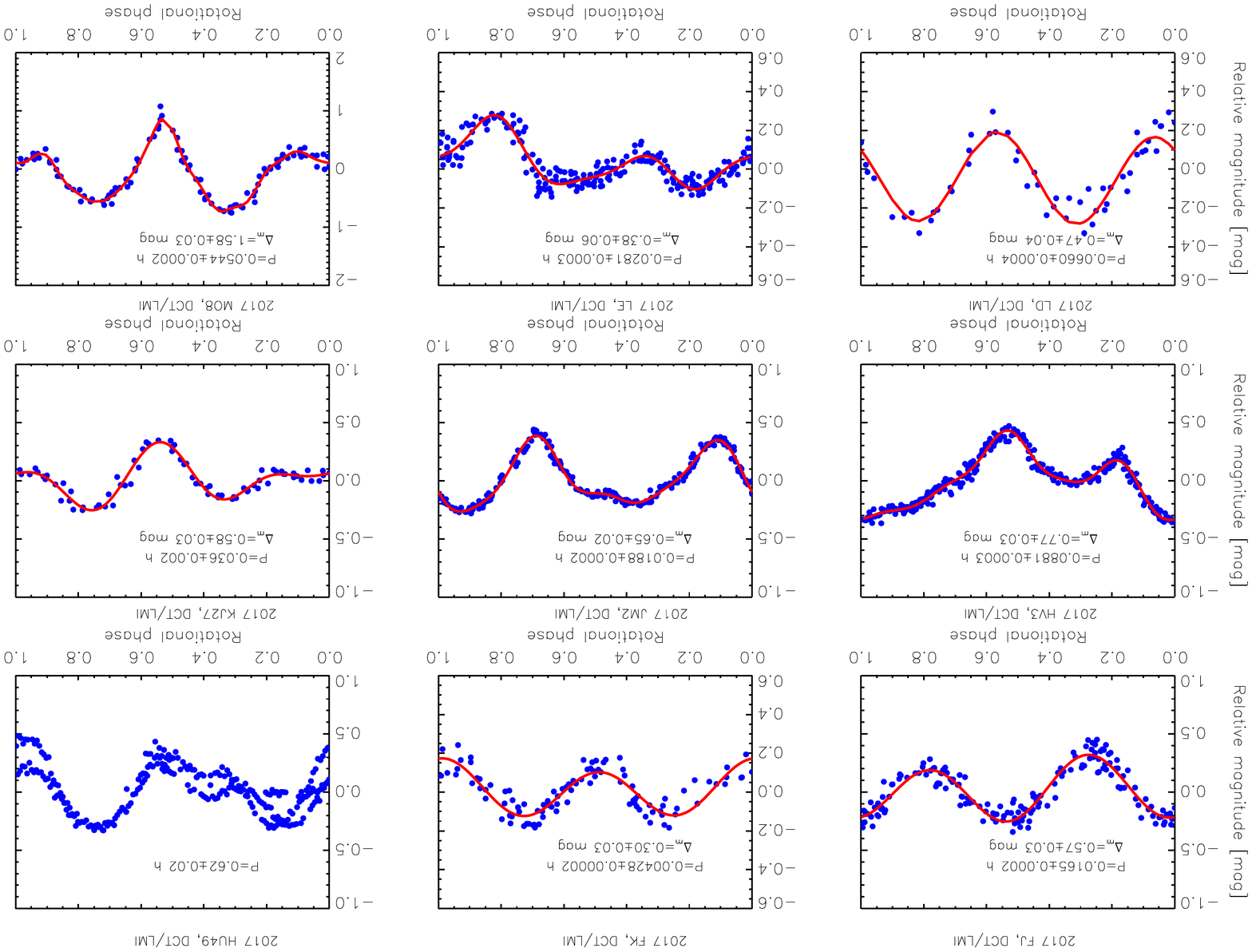} 
\caption {Continued }
\label{fig:LC13}
\end{figure}

\clearpage

\begin{figure}
\includegraphics[width=24cm, angle=90]{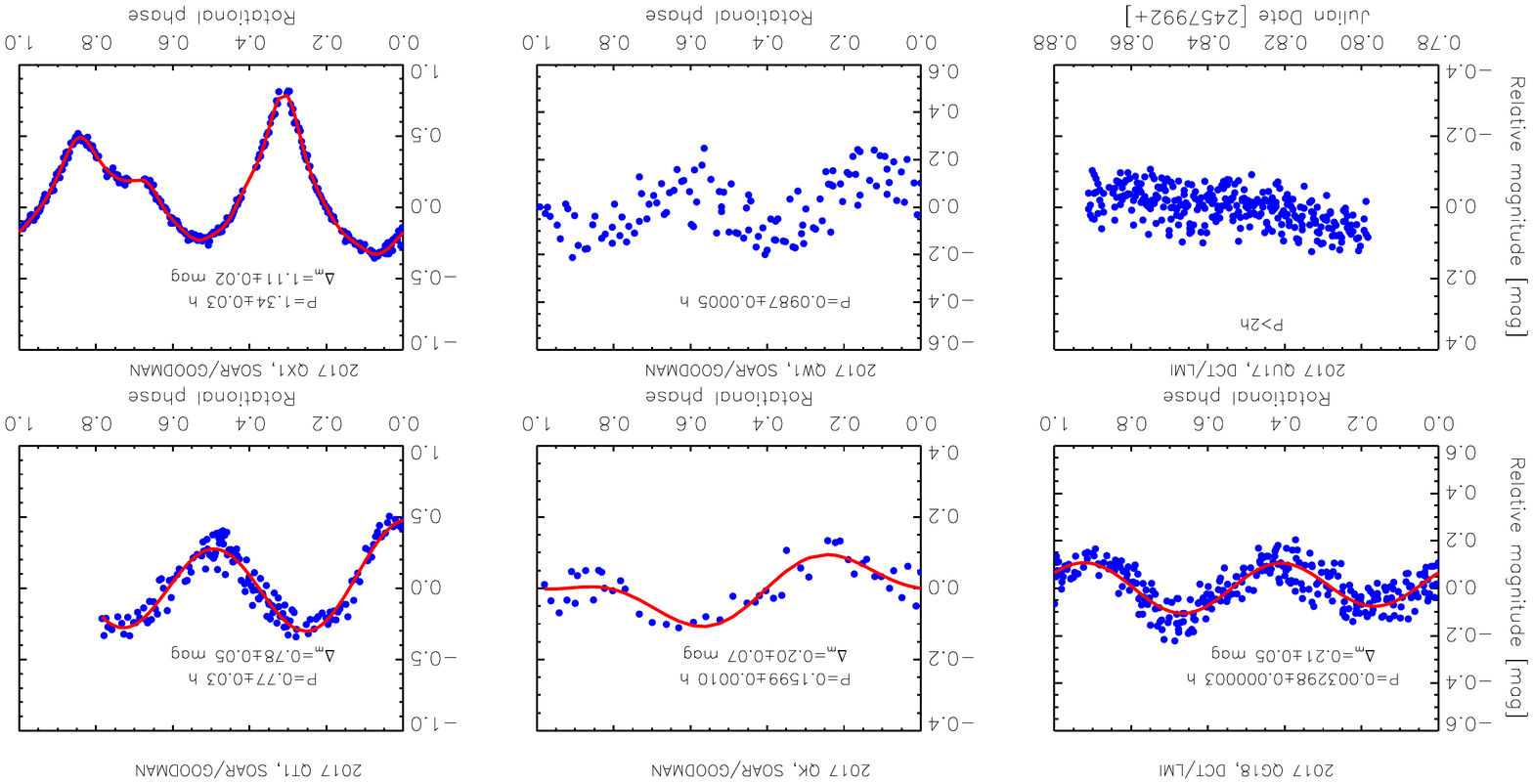} 
\caption {Continued }
\label{fig:LC13}
\end{figure}

\clearpage








\begin{thebibliography}{} 

\bibitem[Abell et 
al.(2009)]{Abell2009} Abell, P.~A., Korsmeyer, D.~J., Landis, R.~R., et al.\ 2009, Meteoritics and Planetary Science, 44, 1825 
\bibitem[Bambach et al.(2018)]{Bambach2018} Bambach, P., Deller, J., Vilenius, E., et al.\ 2018, arXiv:1805.01750 
\bibitem[Benner et al.(2015)]{Benner2015} Benner, L.~A.~M., Busch, M.~W., Giorgini, J.~D., Taylor, P.~A., \& Margot, J.-L.\ 2015, Asteroids IV, 165  
\bibitem[Brozovi{\'c} et al.(2011)]{Brozovic2011} Brozovi{\'c}, M., Benner, L.~A.~M., Taylor, P.~A., et al.\ 2011, \icarus, 216, 241 
\bibitem[Busch et al.(2011)]{Busch2011} Busch, M.~W., Ostro, S.~J., Benner, L.~A.~M., et al.\ 2011, \icarus, 212, 649 
 \bibitem[Carry(2012)]{Carry2012} Carry, B.\ 2012, \planss, 73, 98 
\bibitem[Fujiwara et al.(2006)]{Fujiwara2006} Fujiwara, A., Kawaguchi, J., Yeomans, D.~K., et al.\ 2006, Science, 312, 1330 
\bibitem[Hanu{\v s} et al.(2013)]{Hanus2013} Hanu{\v s}, J., {\v D}urech, J., Bro{\v z}, M., et al.\ 2013, \aap, 551, A67 
\bibitem[Hestroffer et al.(2017)]{Hestroffer2017} Hestroffer, D., Agnan, M., Segret, B., et al.\ 2017, AGU Fall Meeting Abstracts,  
\bibitem[Holsapple(2004)]{Holsapple2004} Holsapple, K.~A.\ 2004, 
\icarus, 172, 272 
\bibitem[Holsapple(2007)]{Holsapple2007} Holsapple, K.~A.\ 2007, 
\icarus, 187, 500 
\bibitem[Kikwaya Eluo(2018)]{Kikwaya2018} Kikwaya Eluo, J.-B.\ 2018, The Vatican Observatory, Castel Gandolfo: 80th Anniversary Celebration, 51, 27 
\bibitem[Kwiatkowski et 
al.(2010)]{Kwiatkowski2010} Kwiatkowski, T., Polinska, M., Loaring, N., et al.\ 2010, \aap, 511, A49 
\bibitem[Lacerda \& Luu(2003)]{Lacerda2003} Lacerda, P., \& Luu, J.\ 2003, \icarus, 161, 174 
\bibitem[La Spina et al.(2004)]{LaSpina2004} La Spina, A., Paolicchi, P., Kryszczy{\'n}ska, A., \& Pravec, P.\ 2004, \nat, 428, 400 
\bibitem[Li et al.(2015)]{Li2015} Li, J.-Y., Helfenstein, P., Buratti, B., Takir, D., \& Clark, B.~E.\ 2015, Asteroids IV, 129 
\bibitem[Margot et al.(2002)]{Margot2002} Margot, J.~L., Nolan, 
M.~C., Benner, L.~A.~M., et al.\ 2002, Science, 296, 1445 
\bibitem[Michalowski \& Velichko(1990)]{Michalowski1990} Michalowski, T., \& Velichko, F.~P.\ 1990, \actaa, 40, 321 
\bibitem[Nolan et al.(2013)]{Nolan2013} Nolan, M.~C., Magri, C., Howell, E.~S., et al.\ 2013, \icarus, 226, 629 
\bibitem[Ostro et al.(2006)]{Ostro2006} Ostro, S.~J., Margot, J.-L., Benner, L.~A.~M., et al.\ 2006, Science, 314, 1276 
\bibitem[Pravec 
\& Harris(2000)]{Pravec2000} Pravec, P., \& Harris, A.~W.\ 2000, \icarus, 148, 12
\bibitem[Pravec et al.(2000)]{Pravec2000b} Pravec, P., Hergenrother, C., Whiteley, R., et al.\ 2000, \icarus, 147, 477  
\bibitem[Pravec et al.(2002)]{Pravec2002} Pravec, P., Harris, 
A.~W., \& Michalowski, T.\ 2002, Asteroids III, 113 
\bibitem[Pravec et al.(2005)]{Pravec2005} Pravec, P., Harris, A.~W., Scheirich, P., et al.\ 2005, \icarus, 173, 108 
\bibitem[Pravec et al.(2006)]{Pravec2006} Pravec, P., Scheirich, P., Ku{\v s}nir{\'a}k, P., et al.\ 2006, \icarus, 181, 63
\bibitem[Pravec \& Harris(2007)]{Pravec2007} Pravec, P., \& Harris, A.~W.\ 2007, \icarus, 190, 250 
\bibitem[Pravec et al.(2008)]{Pravec2008} Pravec, P., Harris, A.~W., Vokrouhlick{\'y}, D., et al.\ 2008, \icarus, 197, 497 
\bibitem[Reddy et al.(2015)]{Reddy2015} Reddy, V., Dunn, T.~L., Thomas, C.~A., Moskovitz, N.~A., \& Burbine, T.~H.\ 2015, Asteroids IV, 43  
\bibitem[Richardson et al.(2005)]{Richardson2005} Richardson, D.~C., 
Elankumaran, P., \& Sanderson, R.~E.\ 2005, \icarus, 173, 349 
\bibitem[Shoemaker 
\& Helin(1978)]{Shoemaker1978} Shoemaker, E.~M., \& Helin, E.~F.\ 1978, Reports of Planetary Geology Program, 20 
\bibitem[Taylor(2009)]{Taylor2009} Taylor, P.~A.\ 2009, Ph.D.~Thesis, Cornell University.  
\bibitem[Thirouin et al.(2016)]{Thirouin2016} Thirouin, A., Moskovitz, N., Binzel, R.~P., et al.\ 2016, \aj, 152, 163 
\bibitem[Veverka et al.(2000)]{Veverka2000} Veverka, J., Robinson, M., Thomas, P., et al.\ 2000, Science, 289, 2088 
\bibitem[Vokrouhlick{\'y} et al.(2015)]{Vokrouhlicky2015} Vokrouhlick{\'y}, D., Bottke, W.~F., Chesley, S.~R., Scheeres, D.~J., \& Statler, T.~S.\ 2015, Asteroids IV, 509 
\bibitem[Warner et al.(2009)]{Warner2009} Warner, B.~D., Harris, 
A.~W., \& Pravec, P.\ 2009, \icarus, 202, 134 
\bibitem[Warner(2014)]{Warner2014} Warner, B.~D.\ 2014, Minor Planet Bulletin, 41, 157 
\bibitem[Warner(2015)]{Warner2015CJ1} Warner, B.~D.\ 2015, Minor Planet Bulletin, 42, 41 
\bibitem[Warner(2015)]{Warner2015} Warner, B.~D.\ 2015, Minor Planet Bulletin, 42, 256 
\bibitem[Warner(2015)]{Warner2015SM143} Warner, B.~D.\ 2015, Minor Planet Bulletin, 42, 115 
\bibitem[Zappala et 
al.(1990)]{Zappala1990} Zappala, V., Cellino, A., Barucci, A.~M., Fulchignoni, M., \& Lupishko, D.~F.\ 1990, \aap, 231, 548 

\end{thebibliography}
\end{document}